\documentclass[10pt]{iopart}
\usepackage{graphicx}
\eqnobysec

\newcommand{\beq}{\begin{equation}}
\newcommand{\beqa}{\begin{eqnarray}}
\newcommand{\eeq}{\end{equation}}
\newcommand{\eeqa}{\end{eqnarray}}
\newcommand{\abs}[1]{\vert#1\vert}
\newcommand{\cm}{{\rm cm}}
\newcommand{\dd}{{\rm d}}
\renewcommand{\dh}{{\Delta_h}}
\newcommand{\dv}{{\Delta_v}}
\newcommand{\eps}{{\varepsilon}}
\newcommand{\goe}{{\rm GOE}}
\newcommand{\gue}{{\rm GUE}}
\newcommand{\hb}{{h_{\rm b}}}
\renewcommand{\max}{{\rm max}}
\newcommand{\mb}{{M_{\rm b}}}
\newcommand{\mean}[1]{\langle#1\rangle}
\newcommand{\s}{{\sigma}}
\newcommand{\sea}{{\rm sea}}
\newcommand{\var}{{\mathop{\rm var}\nolimits}\,}
\renewcommand{\H}{{\cal H}}
\newcommand{\N}{{\rm N}}
\renewcommand{\S}{{\rm S}}
\newcommand{\E}{{\rm E}}
\newcommand{\W}{{\rm W}}

\begin{document}

\title{Generic phase coexistence in the totally asymmetric kinetic Ising model}

\author{Claude Godr\`eche and Jean-Marc Luck}

\address{Institut de Physique Th\'eorique, Universit\'e Paris-Saclay, CEA and CNRS,
91191~Gif-sur-Yvette, France}

\begin{abstract}
The physical analysis of generic phase coexistence in the North-East-Center
Toom model was originally given by Bennett and Grinstein.
The gist of their argument relies on the dynamics of interfaces and droplets.
We revisit the same question for a specific totally asymmetric kinetic
Ising model on the square lattice.
This nonequilibrium model possesses the remarkable property that
its stationary-state measure in the absence of a magnetic field coincides
with that of the usual ferromagnetic Ising model.
We use both analytical arguments and numerical simulations in order to make progress
in the quantitative understanding of the phenomenon of generic phase coexistence.
At zero temperature a mapping onto the TASEP allows an exact determination
of the time-dependent shape of the ballistic interface
sweeping a large square minority droplet of up or down spins.
At finite temperature, measuring the mean lifetime of such a droplet
allows an accurate measurement of its shrinking velocity $v$,
which depends on temperature $T$ and magnetic field $h$.
In the absence of a magnetic field, $v$ vanishes with an exponent
$\Delta_v\approx2.5\pm0.2$ as the critical temperature $T_c$ is approached.
At fixed temperature in the ordered phase, $v$ vanishes at the phase-boundary fields
$\pm h_{\rm b}(T)$ which mark the limits of the coexistence region.
The latter fields vanish with an exponent
$\Delta_h\approx3.2\pm0.3$ as $T_c$ is approached.
\end{abstract}

\ead{\mailto{claude.godreche@ipht.fr,jean-marc.luck@ipht.fr}}

\maketitle

\section{Introduction and summary}

As was demonstrated by Toom~\cite{toom},
an irreversible probabilistic cellular automaton can exhibit generic phase coexistence,
i.e., two stable stationary phases can coexist in a whole domain of parameter space.
This property is alternatively referred to as generic bistability or generic nonergodicity.
Generic phase coexistence in nonequilibrium systems
stands in contrast with the case of equilibrium systems,
where coexistence requires the equality of the thermodynamical free energies
of the two phases, in agreement with Gibbs phase rule,
and therefore some fine tuning of parameters.
Taking the example of the two-dimensional ferromagnetic Ising model
in its low-temperature phase,
the two ordered phases only coexist at zero magnetic field.

A prototypical instance of a nonequilibrium model exhibiting generic bistability
is the two-dimensional North-East-Center (NEC) Toom model,
originally analysed in simple physical terms by Bennett and
Grinstein~\cite{grinstein1}, then further investigated
in~\cite{grinstein2,grinstein3,grinstein4}.
This model consists of a square lattice of Ising spins $\s=\pm1$,
evolving according to a synchronous discrete-time dynamics,
where the spin $\s$ located at a generic site, named the central spin,
is only influenced by its North and East neighbours $\s_\N$ and $\s_\E$.
At each time step all spins are updated synchronously,
according to the following rules:

\begin{itemize}
\item First, a deterministic majority rule is applied: if two or more of the
spins $\s$, $\s_\N$ and $\s_\E$ point up (down), then the value $\s$ of the
central spin at time $t+1$ is set to $+1$ ($-1$).
\item Then, a probabilistic rule is applied: if $\s(t+1)=+1$, then it is
flipped to $-1$ with probability $p$, while if $\s(t+1)=-1$, it is flipped to $+1$
with probability $q$.
\end{itemize}

The corresponding transition probabilities are given in table~\ref{tab:toom}.
The noise parameter $r=p+q$ is analogous to a temperature, while the
symmetry-breaking parameter $h=q-p$ is analogous to a magnetic field.
Indeed, when $q>p$, up spins are favored over down spins.
The remarkable property of this model is that its stationary-state phase diagram
shows coexistence in a whole region of parameter space.
The two stable phases have magnetizations with opposite signs.
As discussed in~\cite{grinstein1,grinstein2,grinstein3,grinstein4},
the same holds true if, instead of being updated synchronously,
the spins are updated in a random sequential manner.
The rate at which the central spin is flipped reads
\beq\label{eq:nec}
w=\frac{1}{2}\left(b-c\,\s(\s_\N+\s_\E)+c\,\s_\N\s_\E-h\s\right),
\eeq
with $b=(1+r)/2$ and $c=(1-r)/2$.
The values given in table~\ref{tab:toom} are now interpreted as rates.

\begin{table}[!ht]
\begin{center}
\begin{tabular}{|c|c|}
\hline
spin-flip move&prob./rate\\
\hline
$+++\;\to\;+-+$&$p$\\
$++-\;\to\;+--$&$p$\\
$-++\;\to\;--+$&$p$\\
$-+-\;\to\;---$&$1-q$\\
$+-+\;\to\;+++$&$1-p$\\
$+--\;\to\;++-$&$q$\\
$--+\;\to\;-++$&$q$\\
$---\;\to\;-+-$&$q$\\
\hline
\end{tabular}
\caption{List of spin-flip moves for the NEC Toom model,
with the convention ($\s_\N,\s,\s_\E$) for local spin configurations,
and expressions of the corresponding transition probabilities (discrete time)
or rates (continuous time).}
\label{tab:toom}
\end{center}
\end{table}

\begin{figure}[!ht]
\begin{center}
\includegraphics[angle=-90,width=.4\linewidth]{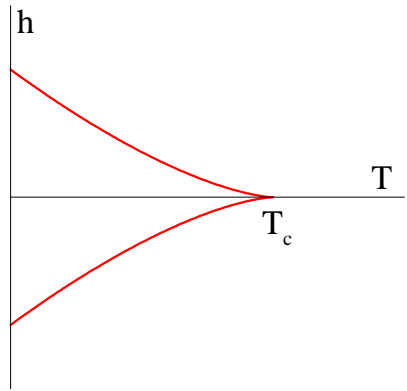}
\caption{\small
Schematic phase diagram of a spin model exhibiting generic nonergodicity
in the plane of temperature $T$ and magnetic field~$h$.
The coexistence region is delimited by curves of the form $h=\pm\hb(T)$.}
\label{phase}
\end{center}
\end{figure}

The gist of the phenomenon of generic bistability relies on the dynamics of interfaces.
Let us summarize the argument put forward
originally in~\cite{grinstein1}, then further expanded
in~\cite{grinstein2,grinstein3,grinstein4}.
Consider a ferromagnetic system of Ising spins
in its low-temperature phase in the absence of a magnetic field.
If the microscopic dynamical rules are reversible,
a flat interface will not move on average,
and a curved one will move only under the effect of surface tension.
For dynamical rules such as those of the NEC Toom model,
which owe their irreversibility to a spatial anisotropy,
a flat interface generically moves ballistically
with a velocity depending on the orientation of the interface.
In the presence of a weak magnetic field~$h$,
this velocity will be corrected by a term proportional to $h$,
and therefore remains nonzero in a whole domain of parameter space.
This picture concerning flat interfaces
can now be completed by an argument concerning droplet dynamics.
If an ordered phase contains a large droplet of the other phase,
this droplet will shrink ballistically,
i.e., in a time proportional to its diameter,
as a consequence of the orientation dependence of the interface velocity.
So, such a dynamics is able to spontaneously heal defects in both ordered phases,
with large defects shrinking on a ballistic time scale.
The simultaneous stability of the two ordered phases
takes place in a whole coexistence region,
where both noise (measured by the temperature $T$)
and bias (measured by the magnetic field $h$) are small enough.
This coexistence region is shown schematically in figure~\ref{phase}.
It is delimited by curves of the form $h=\pm\hb(T)$,
where the phase-boundary field vanishes according to
the power law $\hb(T)\sim(T_c-T)^\dh$
as the critical temperature is approached.

In addition to the NEC Toom model, other models exhibiting generic bistability
were subsequently analysed by several groups, mostly in the two-dimensional
setting~\cite{heringa,lebo1,gls,hinrich,munoz,muk,gje},
but also in three~\cite{h3} and even higher~\cite{wle} dimensions.
Besides its major conceptual role recalled above,
the ballistic shrinking of large droplets (or islands) of prescribed form
has been instrumental in the numerical determination of the boundaries
of the nonergodic region
performed in~\cite{grinstein1,grinstein3,heringa,gje}.
Let us mention a review putting the phenomenon of generic coexistence
in a very broad perspective~\cite{mckay},
as well as a very recent work~\cite{rd} pinpointing
various subtleties related to spatial aspects of phase coexistence
in nonequilibrium systems.

In the present work we revisit the investigations made
in~\cite{grinstein1,grinstein2,grinstein3,heringa}, with the aim of giving a
more quantitative analysis of generic bistability in two-dimensional spin systems,
putting a special emphasis both on the zero-temperature limit
and on the vicinity of the critical point.
To this end we focus on a special model in the class of totally asymmetric NEC dynamics.
This model, introduced initially by K\"unsch~\cite{kunsch} in a slightly
different form (see~(\ref{eq:kunsch}) below),
possesses the remarkable property~\cite{kunsch,gb2009,cg2013}
of having the same stationary measure as the two-dimensional Ising model
at temperature $T$ in the absence of a magnetic field,
with the usual nearest-neighbour ferromagnetic Hamiltonian
\beq\label{eq:energy}
\H=-J\sum_{\langle i,j\rangle}\s_i\s_j,
\eeq
where the sum runs over pairs $\langle i,j\rangle$ of neighbouring spins.
The rate defining the model reads
\beqa
w
&=&\frac{\alpha}{2}\Bigl(1-\gamma\s(\s_E+\s_N)+\gamma^2\,\s_E\s_N\Bigr)
\nonumber\\
&=&\frac{\alpha}{2}(1-\gamma\s\s_E)(1-\gamma\s\s_N),
\label{eq:kun}
\eeqa
where $\alpha$ is an arbitrary frequency fixing the time scale, $\s$ is the flipping spin,
$\s_\N$ and~$\s_\E$ are the North and East neighbouring spins, and
\beq
\gamma=\tanh 2K,
\eeq
where $K=J/T$ is the reduced coupling constant.
This model is the totally asymmetric version of the two-dimensional
{\it directed Ising model}, investigated in~\cite{gb2009,gp2014,gp2015},
whose definition encompasses the case where the dynamics is partially asymmetric,
i.e., where the South and West spins keep an influence on the central spin.
A number of physical properties of this model have been described in these references.
In the one-dimensional situation too, the general directed Ising model
has rich dynamical properties~\cite{cg2011,gl2015}.

The expression~(\ref{eq:kun}) is invariant under up-down spin symmetry and
satisfies the condition of global balance,
that is to say, leads to a Gibbsian stationary measure with respect to the
Hamiltonian~(\ref{eq:energy}).
The dynamics thus defined is irreversible, since the condition of detailed
balance is not satisfied~\cite{kunsch,gb2009,cg2013}.
This model is actually the {\it unique} Gibbsian
two-dimensional spin system with totally asymmetric NEC dynamics.
Imposing the condition of global balance with respect to the
Hamiltonian~(\ref{eq:energy}) on a NEC dynamics indeed uniquely determines
the rate~(\ref{eq:kun})~\cite{gb2009,cg2013}.
Fixing the time scale in~(\ref{eq:kun}) by the choice $\alpha=2\cosh^2 2K$
yields the exponential form $w=\exp(-2K\s(\s_\N+\s_\E))$.
Note that the expression of the rate originally given by K\"unsch~\cite{kunsch},
\beq\label{eq:kunsch}
w=\e^{-K\s(\s_\N+\s_\E)},
\eeq
leads to a ferromagnetic Gibbs measure at temperature~$T/2$.

At finite temperature and in the presence of an external magnetic field $h$,
it can be shown, along the lines of~\cite{cg2013}, that
there is no NEC dynamics yielding global balance with respect to the full Hamiltonian
\beq
\H=-J\sum_{\langle i,j\rangle}\s_i\s_j-h\sum_i\s_i.
\label{hfull}
\eeq
In this work we choose to introduce the effect of the magnetic field $h$
in a multiplicative form,
as originally done by Glauber for the one-dimensional kinetic Ising model~\cite{glauber}.
We shall thus consider the dynamics defined by the rate function
\beq\label{eq:wfield}
w=\frac{\alpha}{2}(1-\gamma\s\s_\N)(1-\gamma\s\s_\E)(1-\kappa\s),
\eeq
with
\beq
\kappa=\tanh(h/T).
\eeq
The values taken by this rate function for the eight possible local
configurations $(\s_N,\s,\s_E)$ are given in table~\ref{tab:moves}.
As analysed in detail in the present work, this model exhibits generic phase coexistence.
Though its stationary measure is not known in closed form
for generic values of the parameters $\gamma$ and $\kappa$,
it is however known exactly in two special cases.
In the absence of a magnetic field ($\kappa=0$), as said above,
the stationary measure is the same as that of the equilibrium model with
Hamiltonian~(\ref{eq:energy}),
which is~(\ref{hfull}) restricted to its first (ferromagnetic) term.
In the absence of interactions ($\gamma=0$),
the dynamics obeys detailed balance with respect to the second (paramagnetic) term
of the Hamiltonian~(\ref{hfull}).
These special cases demonstrate that temperature $T$ and magnetic field $h$
are natural and properly normalized extensions
of what they are in the usual Ising model.

\begin{table}[!ht]
\begin{center}
\begin{tabular}{|c|c|}
\hline
spin-flip move&rate\\
\hline
$+++\;\to\;+-+$&$\frac{\alpha}{2}(1-\gamma)^2(1-\kappa)$\\
$++-\;\to\;+--$&$\frac{\alpha}{2}(1-\gamma^2)(1-\kappa)$\\
$-++\;\to\;--+$&$\frac{\alpha}{2}(1-\gamma^2)(1-\kappa)$\\
$-+-\;\to\;---$&$\frac{\alpha}{2}(1+\gamma)^2(1-\kappa)$\\
$+-+\;\to\;+++$&$\frac{\alpha}{2}(1+\gamma)^2(1+\kappa)$\\
$+--\;\to\;++-$&$\frac{\alpha}{2}(1-\gamma^2)(1+\kappa)$\\
$--+\;\to\;-++$&$\frac{\alpha}{2}(1-\gamma^2)(1+\kappa)$\\
$---\;\to\;-+-$&$\frac{\alpha}{2}(1-\gamma)^2(1+\kappa)$\\
\hline
\end{tabular}
\caption{List of spin-flip moves for the totally asymmetric kinetic Ising
model,
with the convention ($\s_\N,\s,\s_\E$) for local spin configurations,
and expressions of the corresponding rates~(\ref{eq:wfield}).}
\label{tab:moves}
\end{center}
\end{table}

The main motivation behind our choice of the rate~(\ref{eq:wfield})
is that it is more advantageous,
especially from the standpoint of numerical simulations,
to study generic bistability in a model
whose stationary properties are known exactly in the absence of a magnetic field.
In particular, the critical temperature is that of the two-dimensional
ferromagnetic Ising model, given by $\sinh 2K_c=1$.
Henceforth we choose energy units such that $J=1$, and so $K=1/T$.
We have thus in particular
\beq
T_c=\frac{2}{\ln(1+\sqrt{2})}\approx 2.269.
\eeq

Coming back to the NEC Toom model~(\ref{eq:nec}), let us point out that this model
is just a generic member of the class of NEC dynamics, with no specific property.
In particular its stationary-state measure is not known in closed form,
even in the absence of a bias ($p=q$).
Let us note however that the rate~(\ref{eq:nec}) with $p=q=0$ is
the same as~(\ref{eq:wfield}) with $\gamma=1$, $h=0$ and $\alpha=1/2$.
This can also be checked by comparing tables~\ref{tab:toom}
and~\ref{tab:moves}.

Then, a natural question to ask is: what are the minimal conditions
required to produce generic nonergodicity?
It was argued in~\cite{grinstein3}
that {\it the only indispensable ingredient seems to
be irreversibility in the specific form of spatially asymmetric rules}.
Restricting ourselves to spin models,
this statement is confirmed by the behaviour of the NEC Toom model,
of the model studied in~\cite{heringa},
and of the directed kinetic Ising model investigated in the present work.
There are however exceptions to this picture.
One notable case is the following one.
Starting with the usual rate function
\beq\label{glau2D}
w=\frac{\alpha}{2}\big(1-\s\tanh[K(\s_\N+\s_\E+\s_\W+\s_\S)]\big),
\eeq
which is the direct generalisation to two dimensions of the Glauber rate
function~\cite{glauber},
and just suppressing the influence of the West and South spins,
results in the following expression, after reduction on a basis of spin operators:
\beq\label{eq:voter}
w=\frac{\alpha}{2}\left(1-\frac{\gamma}{2}\s(\s_\N+\s_\E)\right).
\eeq
This rate function describes a truncated noisy voter model
with no West and South influence~\cite{gb2009,cg2013}.
Just as the isotropic two-dimensional voter model,
this model has no phase transition~\cite{gb2009,lima}.

We now give an overview and a discussion of the main results obtained in the present work.
In section~\ref{sec:zerot} we study the zero-temperature limit of the model,
allowing the presence of a vanishingly small magnetic field, proportional to temperature.
The only flippable spins are located in the NE corners of interfaces
separating domains of up and down spins (see figure~\ref{fig:interface}).
As a consequence, a descending interface moves ballistically,
while an ascending one is immobile.
A mapping onto the totally asymmetric exclusion process (TASEP) allows
to determine the angle-dependent velocity $v(\theta)$ of a flat interface.
This velocity takes its maximal value $v$
when the interface is normal to the bias of the dynamics,
which is oriented from NE to SW.
In the absence of a magnetic field, we have $v=v_0=\alpha/\sqrt{2}$
(section~\ref{sec:mapping}).
We then focus our attention on the fate
of a large square droplet of linear size $N$, embedded in a sea of opposite spins.
This droplet is swept by a ballistic interface propagating at velocity~$v$ in
the direction of the bias.
Accordingly, its lifetime $t^\star$ grows proportionally to $N$
(section~\ref{sec:lifetime}).
This interface is however not straight, even if the droplet is very large.
Its time-dependent shape is obtained by means of the hydrodynamic approach to the TASEP,
based on the inviscid Burgers equation (section~\ref{sec:interface}).
It is given by an arc of parabola (see figure~\ref{fig:carre}).
We finally consider the fluctuations of the droplet's lifetime $t^\star$
around the ballistic law.
According to the KPZ theory, these fluctuations scale as $N^{1/3}$ times
a dimensionless random variable $\chi$ with a non-trivial distribution
(section~\ref{sec:fluctuat}).
Numerical simulations strongly corroborate the ansatz that~$\chi$
is proportional to the variable $\xi_\gue$
distributed according to the Tracy-Widom~$F_2$ law (see figure~\ref{fig:hchi}).

In section~\ref{sec:finitet}
we investigate the model at finite temperature and zero field by means of
numerical simulations.
The dynamics of a large droplet concomitantly exhibits two distinct features:
its center of mass is advected ballistically
in the direction of the bias with velocity $v_\cm$,
whereas the droplet shrinks ballistically with velocity $v$,
and therefore disappears after a time $t^\star$ which grows linearly with $N$.
As temperature increases, $v_\cm$ increases steadily,
while the droplet becomes progressively rounder,
and its lifetime $t^\star$ increases very fast as $T_c$ is approached
(section~\ref{sec:overall}).
The quantitative determination of the
shrinking velocity $v$ relies on a precise definition of $t^\star$.
In analogy with the zero-temperature situation,
where $t^\star$ is the time where the system reaches its ground state,
in the finite-temperature situation we define the stopping time $t^\star$
as the first time where the mean magnetization exceeds
the equilibrium spontaneous magnetization
(section~\ref{sec:finitev}).
The outcome of this study is shown in figure~\ref{fig:vitesse},
depicting the ratio $v/v_0$ as a function of $T/T_c$.
At low temperature, this ratio departs from unity as $\e^{-2/T}$.
At criticality, a finite-size scaling analysis yields $v\sim(T_c-T)^\dv$,
with $\dv\approx 2.5\pm0.2$
(section~\ref{sec:vcrit}).

Section~\ref{sec:finiteth} is devoted to a numerical investigation
of the generic situation
where the temperature $T$ and the magnetic field~$h$ are both non-zero.
The system exhibits generic bistability in a whole coexistence region of parameter space,
delimited by the phase-boundary fields $\pm\hb(T)$.
For small enough values of the magnetic field,
the situation is similar to what happens for $h=0$,
i.e., a large droplet of down spins disappears in a ballistic time.
The corresponding shrinking velocity $v$ depends on both $T$ and~$h$,
vanishing as $h\to-\hb(T)$, where the up phase loses its stability
(section~\ref{sec:hover}).
An efficient numerical determination of the phase-boundary field
again requires an operational definition of a stopping time $t^{\star}$.
We have chosen to define it as the first time
where the magnetization of a system with droplet initial condition
exceeds that of an initially homogeneous system.
We have also measured the splitting probability $p^\star$ of the up phase,
i.e., the probability for the system with droplet initial condition to end
in the up phase (section~\ref{sec:bdy}).
The finite-size scaling analysis of $p^\star$ yields the most accurate
determination of the phase-boundary field.
The outcome of this study is shown in figure~\ref{fig:h0},
depicting $\hb(T)$ as a function of $T/T_c$.
This figure therefore provides a quantitative version
of the phase diagram shown schematically in figure~\ref{phase}.
At low temperature we have $\hb\approx2-BT$, with $B\approx1.63$,
whereas a critical finite-size scaling analysis yields $\hb\sim(T_c-T)^\dh$,
with $\dh\approx3.2\pm0.3$.
Our predictions for the critical exponents~$\dv$ and $\dh$
are put in perspective with existing analytical and numerical results.
We obtain in particular the estimate $y_i\approx-1.3\pm0.2$
for the irrelevant scaling dimension entering the heuristic approach
of~\cite{heringa}
(section~\ref{sec:hcrit}).
Finally, the magnetization curves of the up phase are shown in
figure~\ref{fig:aim}
for three temperatures in the ordered phase.
These curves are observed to be very regular near the phase-boundary field
(section~\ref{sec:mag}).

\section{Zero-temperature dynamics}
\label{sec:zerot}

We start by considering the zero-temperature limit of the dynamics
($K=1/T\to\infty$, and so $\gamma\to1$).
The parameter $\kappa=\tanh(h/T)$ is kept fixed, thus allowing to consider
vanishingly small magnetic fields $h$, proportional to temperature $T$.

For $\gamma=1$ and $\kappa$ arbitrary, the rate function~(\ref{eq:wfield}) reads
\beq
w=\frac{\alpha}{2}(1-\s\s_\N)(1-\s\s_\E)(1-\kappa\s).
\label{wzero}
\eeq
It takes non-zero values for only two configurations,
corresponding to lines 4 and 5 of table~\ref{tab:moves}:

\begin{itemize}
\item If $\s_\N=\s_\E=-1$, $\s$ is flipped from $+1$ to $-1$ with rate
\beq
w=2\alpha(1-\kappa);
\label{wp}
\eeq
\item If $\s_\N=\s_\E=+1$, $\s$ is flipped from $-1$ to $+1$ with rate
\beq
w=2\alpha(1+\kappa).
\label{wm}
\eeq
\end{itemize}

\begin{figure}[!ht]
\begin{center}
\includegraphics[angle=-90,width=.4\linewidth]{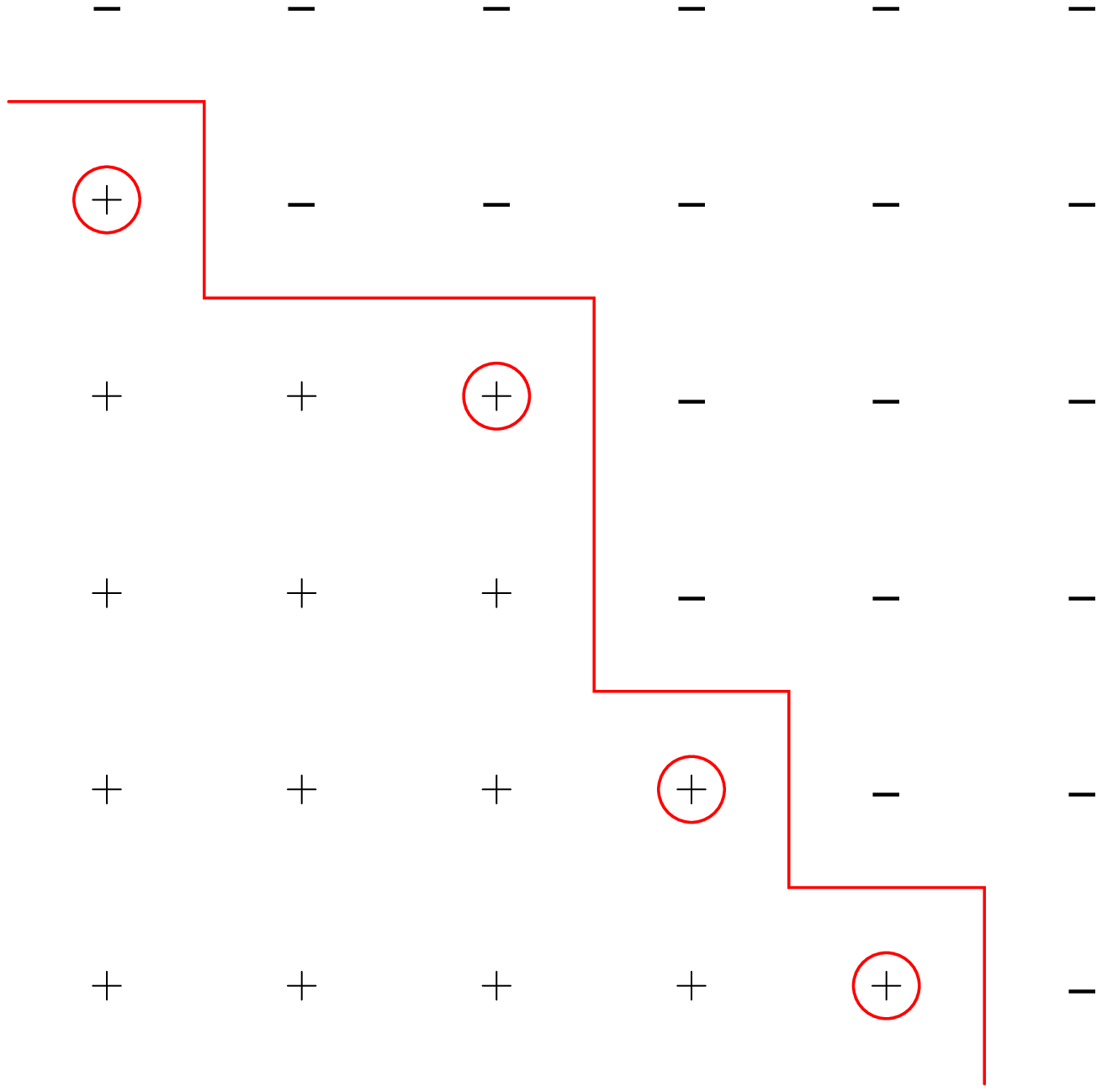}
\caption{\small
A descending interface separating domains of up and down spins.
Flippable spins are circled.
The same picture holds if the roles of $+$ and $-$ spins are exchanged.}
\label{fig:interface}
\end{center}
\end{figure}

The only flippable spins are those located in the NE corners
of interfaces between domains of up and down spins
(see figure~\ref{fig:interface}).
As a consequence, a descending interface is mobile, while an ascending one is not.

\subsection{Mapping onto the TASEP}
\label{sec:mapping}

The dynamics of a descending interface can be exactly mapped onto that
of the totally asymmetric exclusion process (TASEP)~\cite{schutz,liggett}.
To this end, horizontal segments of the interface are identified with particles
(denoted by 1) and vertical ones with holes (denoted by 0).
Flipping a spin corresponds to the displacement of a particle to the right, according to
\beq
10\to01,
\label{1001}
\eeq
with rate
\beq
w=2\alpha(1\pm\kappa),
\eeq
where the plus (minus) sign applies if up spins are above (below) the interface.

The mapping onto the TASEP provides a useful tool to investigate the dynamics.
For the time being, we consider a single macroscopically flat descending interface,
making an angle~$\theta$ with the horizontal ($0<\theta<\pi/2$).
The particle density of the associated TASEP
is equal to the fraction of horizontal segments, i.e.,
\beq
\rho(\theta)=\frac{\cos\theta}{\sin\theta+\cos\theta}.
\eeq
The current across the TASEP in its stationary state reads
$j(\theta)=w\rho(\theta)(1-\rho(\theta))$.
Accordingly, the interface moves with a constant normal velocity~$v(\theta)$,
that can be derived as follows.
A useful intermediate quantity is the time $\tau$ it takes
for each particle to move on average one step forward,
so that the integrated current over a time lapse $\tau$ equals the particle density.
This yields $\tau=\rho(\theta)/j(\theta)$.
During the same time, the interface moves by a distance $d=\cos\theta$
(measured perpendicularly to the direction of the interface)~\cite{gp2015}.
This yields $\tau=d/v(\theta)$.
Putting everything together, we obtain the expression
\beq
v(\theta)=w\frac{\sin\theta\cos\theta}{\sin\theta+\cos\theta}
\eeq
for the angle-dependent normal velocity of a flat descending interface.
This velocity takes its maximal value,
\beq
v=\frac{w}{2\sqrt{2}},
\label{vmax}
\eeq
for $\theta=\pi/4$, i.e., when the interface is normal to the bias of the dynamics,
and moves in the direction of this bias, i.e., from NE to SW.

\subsection{Lifetime of a large square droplet}
\label{sec:lifetime}

Along the lines of earlier works~\cite{grinstein1,grinstein3,heringa,gje,gp2015},
we shall focus our attention on the fate
of a single large droplet embedded in a sea of the opposite phase.
For definiteness and simplicity, we shall consider square droplets.
Other shapes could be considered as well (see~\cite{gp2015}).
In this section, devoted to an analytical study of zero-temperature dynamics,
we consider a large square droplet of linear size~$N$,
initially consisting of up spins,
immersed in an infinite sea of down spins, or vice versa.
In sections~\ref{sec:finitet} and~\ref{sec:finiteth},
devoted to numerical simulations,
the sea itself will be modelled as a larger square of size~$N_\sea$,
with periodic boundary conditions.

As soon as it is prepared and subjected to zero-temperature dynamics,
the droplet is swept by an interface propagating
at the maximal velocity $v$ given by~(\ref{vmax})
along the direction of the bias, from its NE corner to its SW corner.
The distance between these two points being $N\sqrt{2}$,
the lifetime $t^\star$ of a large droplet is given by the ballistic formula
\beq
t^\star\approx\frac{N\sqrt{2}}{v},
\label{tsgen}
\eeq
i.e.,
\beq
t^\star\approx\frac{4N}{w}.
\label{tstar}
\eeq
In the absence of a magnetic field ($\kappa=0$) we have $w=2\alpha$,
both for up and down droplets,
and so the velocity (\ref{vmax}) and the droplet lifetime (\ref{tstar}),
denoted by $v_0$ and~$t^\star_0$, read
\beq
v_0=\frac{\alpha}{\sqrt{2}},\qquad
t^\star_0\approx\frac{2N}{\alpha}.
\label{v0}
\eeq
In the presence of a magnetic field, we have
\beq\label{eq:vtstar}
v=v_0(1\pm\kappa),\qquad
t^\star=\frac{t^\star_0}{1\pm\kappa}.
\eeq
Plus (minus) signs hold for droplets of down (up) spins,
whose lifetime decreases (increases) as a function of the magnetic field.
The lifetimes of the two types of droplets are nevertheless finite
for all values of the reduced field $h/T$,
i.e., for all parameters $\kappa$ in the range $-1<\kappa<1$.

The simplicity of the ballistic formula~(\ref{tstar})
however hides many interesting features of zero-temperature dynamics,
to be described in the sequel:

\begin{itemize}
\item
The ballistic interface is not straight, even if the droplet is very large.
The time-dependent shape of the interface will be derived in section~\ref{sec:interface}
for a large square droplet, using a hydrodynamic approach for the TASEP
based on the Burgers equation.
\item
The hydrodynamic approach can be applied to large droplets with arbitrary
initial shapes.
Ballistic laws involving simple geometric factors however only hold for a class
of convex droplet shapes which are {\it symmetric}
with respect to the direction of the bias,
such as those considered in~\cite{gp2015}: square, circle, right isosceles triangle.
(The droplet depicted schematically in figure 12 of this reference should be
more correctly represented as convex and symmetric.)
For more general shapes,
the hydrodynamic approach yields more involved expressions for the lifetime
$t^\star$.
Skipping details, let us mention that it yields the prediction
\beq
wt^\star\approx\bigl(\sqrt{M}+\sqrt{N}\bigr)^2=M+N+2\sqrt{MN}
\eeq
for the lifetime of a large rectangular droplet of size $M\times N$.
\item
The lifetime $t^\star$ of an individual droplet is a random quantity
which fluctuates around the ballistic law~(\ref{tstar}).
The size of these fluctuations scales as~$N^{1/3}$, according to the KPZ theory
(see section~\ref{sec:fluctuat}).
\end{itemize}

\subsection{Shape of the interface sweeping a large square droplet}
\label{sec:interface}

In this section we give an analytical description
of the time-dependent shape of the interface
sweeping a large square droplet of size $N$,
using the exact mapping of the zero-temperature
dynamics onto the TASEP, as described above.
Throughout sections~\ref{sec:interface} and~\ref{sec:fluctuat},
the time unit is fixed by setting $w=1$.

A square droplet of size $N$ corresponds to the TASEP on a closed interval of length $2N$,
whose initial configuration consists of~$N$ particles followed by $N$ holes.
The dynamics brings the system to the absorbing configuration
where the particles have moved to the right of the holes:
\beq
\underbrace{1\dots1}_{N}\underbrace{0\dots0}_{N}
\longrightarrow
\underbrace{0\dots0}_{N}\underbrace{1\dots1}_{N}.
\label{reac}
\eeq
The lifetime $t^\star$ of the droplet is the random duration of this process.

On the ballistic scale,
i.e., when spatial distance and time are large and proportional,
the appropriate level of description is the hydrodynamic one,
which consists in introducing the coarse-grained particle density $\rho(x,t)$,
which obeys the inviscid Burgers equation~\cite{liggett,rost,lps,dls}
\beq
\frac{\partial\rho}{\partial t}+(1-2\rho)\frac{\partial\rho}{\partial x}=0.
\label{b}
\eeq
The initial condition ($N$ particles followed by $N$ holes) corresponds to a
step for the density:
\beq
\rho(x,0)=\left\{\matrix{
1\quad& (0<x<N),\hfill\cr
0\hfill& (N<x<2N).\hfill
}\right.
\label{bin}
\eeq
The time-dependent density profile $\rho(x,t)$ evolved from this initial condition
yields a parametric representation of the interface sweeping the droplet in the spin model:
\beq
X=\int_0^x\rho(x',t)\,\dd x',\qquad Y=N-x+X.
\label{XYinterface}
\eeq
The solution to the Burgers equation~(\ref{b})
with initial condition~(\ref{bin}) involves two successive stages.

\subsubsection*{Stage~I ($\,0<t<N$).}

\begin{figure}[!ht]
\begin{center}
\includegraphics[angle=-90,width=.445\linewidth]{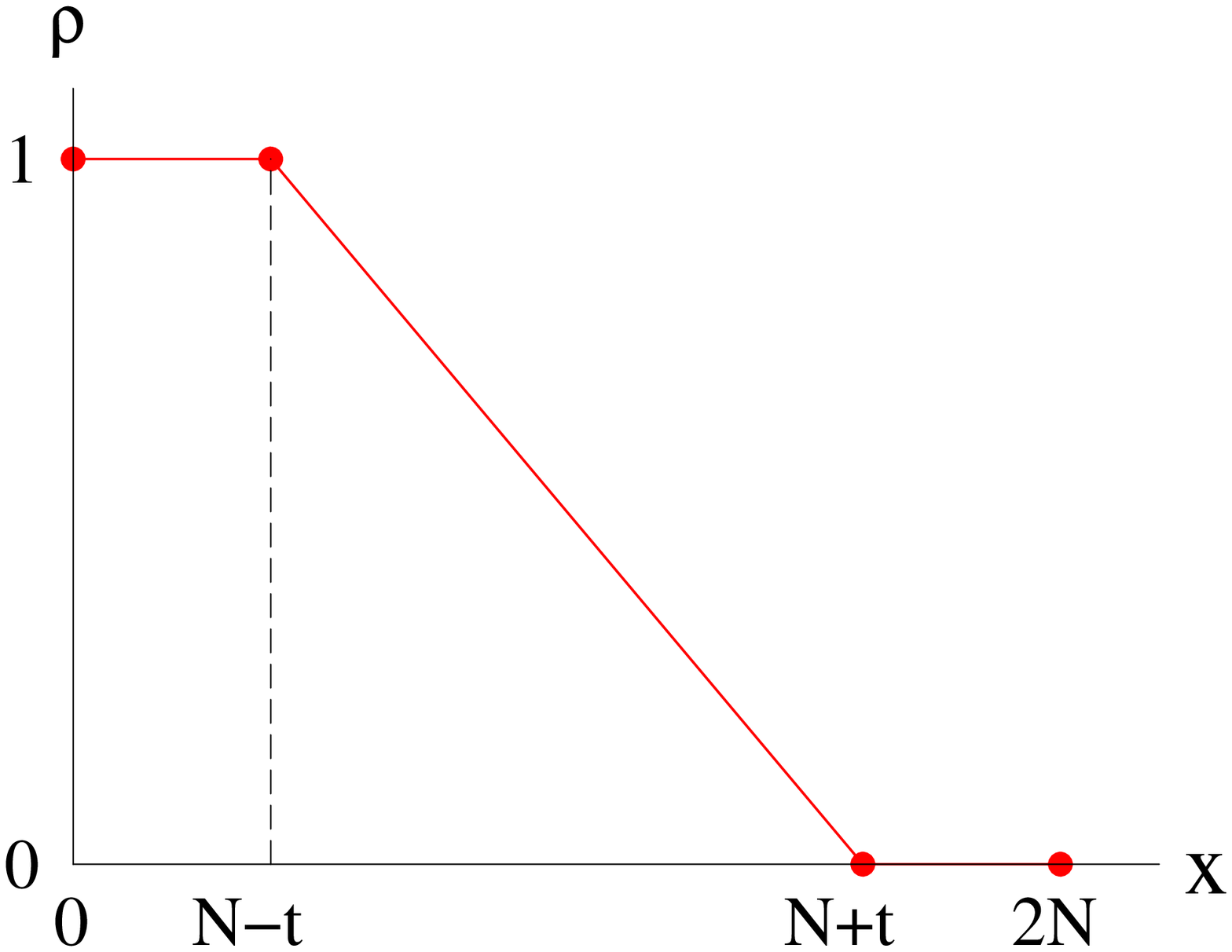}
\hskip 10pt
\includegraphics[angle=-90,width=.355\linewidth]{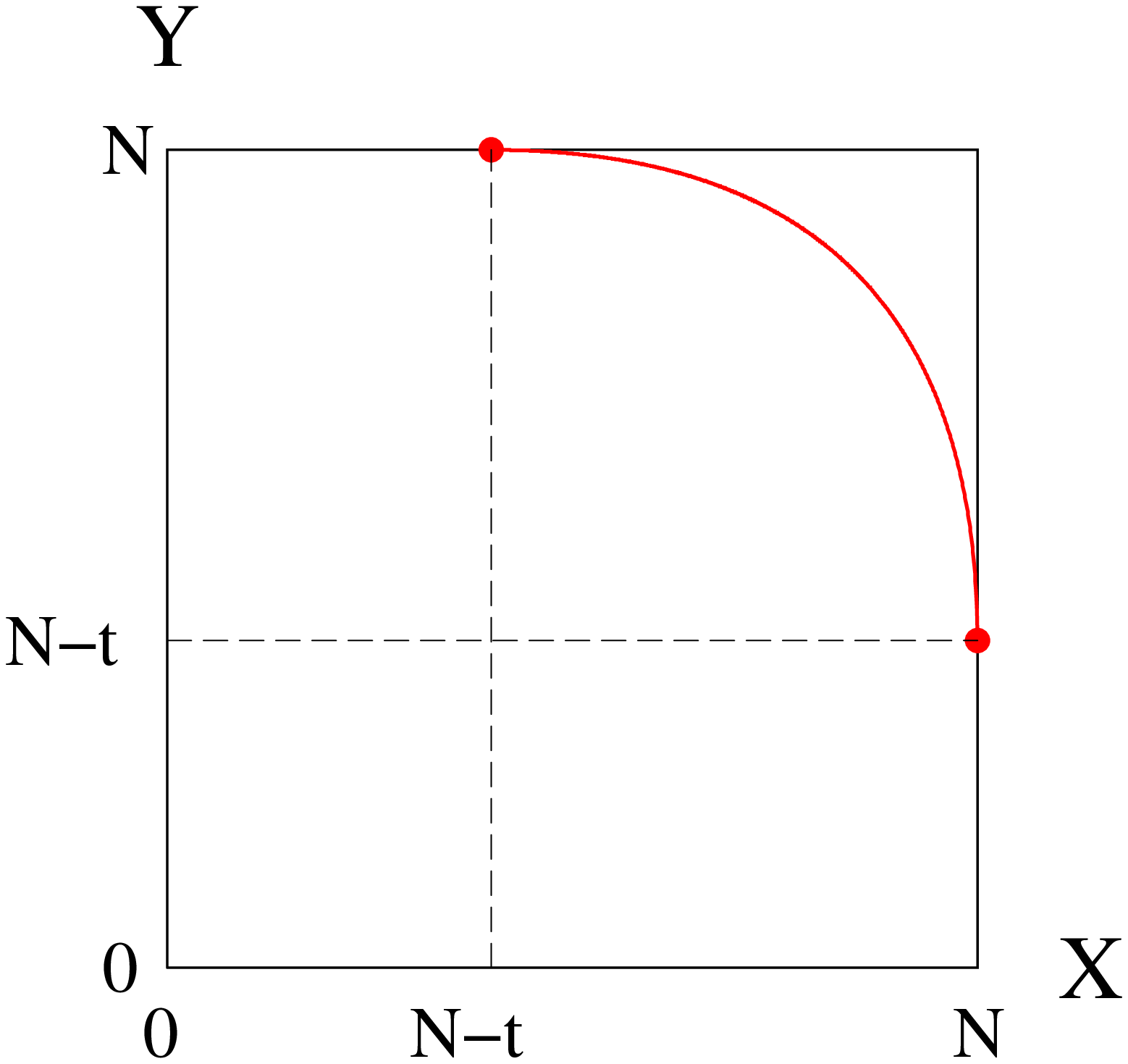}
\caption{\small
Stage I of zero-temperature dynamics for a square droplet.
Left: density profile of the TASEP.
Right: interface sweeping the droplet.}
\label{fig:s1}
\end{center}
\end{figure}

In this first stage, the density profile evolves as if the system were infinite.
The system is thus equivalent to an infinite TASEP with a step initial condition,
where all the sites to the left of an `origin' (here, $x=N$) are occupied,
while all the sites to its right are empty.
In the language of the spin model,
the initial configuration consists of an infinite quadrant of up spins
surrounded by down spins, or vice versa.
The corresponding solution to the Burgers equation
is the so-called rarefaction (or fan) scaling solution~\cite{krbbook}
\beq
\rho(x,t)=\left\{\matrix{
1\hfill&(0<x<N-t),\hfill\cr
(N+t-x)/(2t)\quad&(N-t<x<N+t),\hfill\cr
0\hfill&(N+t<x<2N)\hfill
}\right.
\eeq
(see figure~\ref{fig:s1}, left).
According to~(\ref{XYinterface}),
the corresponding interface is an arc of the parabola with equation
\beq
(X-Y)^2+2t(X+Y-2N)+t^2=0,
\eeq
or, equivalently,
\beq
\sqrt{N-X}+\sqrt{N-Y}=\sqrt{t},
\label{para}
\eeq
a result first derived by Rost~\cite{rost}.
The interface is tangent to the North and East sides of the droplet
(see figure~\ref{fig:s1}, right).
Its apex moves at a constant velocity along the diagonal, according to
\beq
X=Y=N-\frac{t}{4}.
\label{apex}
\eeq
At the end of the first stage $(t=N$),
the density profile is linear over the whole system: $\rho(x,N)=1-x/(2N)$,
while the interface touches the NW and SE corners of the droplet.

\subsubsection*{Stage~II ($N<t<4N$).}

\begin{figure}[!ht]
\begin{center}
\includegraphics[angle=-90,width=.46\linewidth]{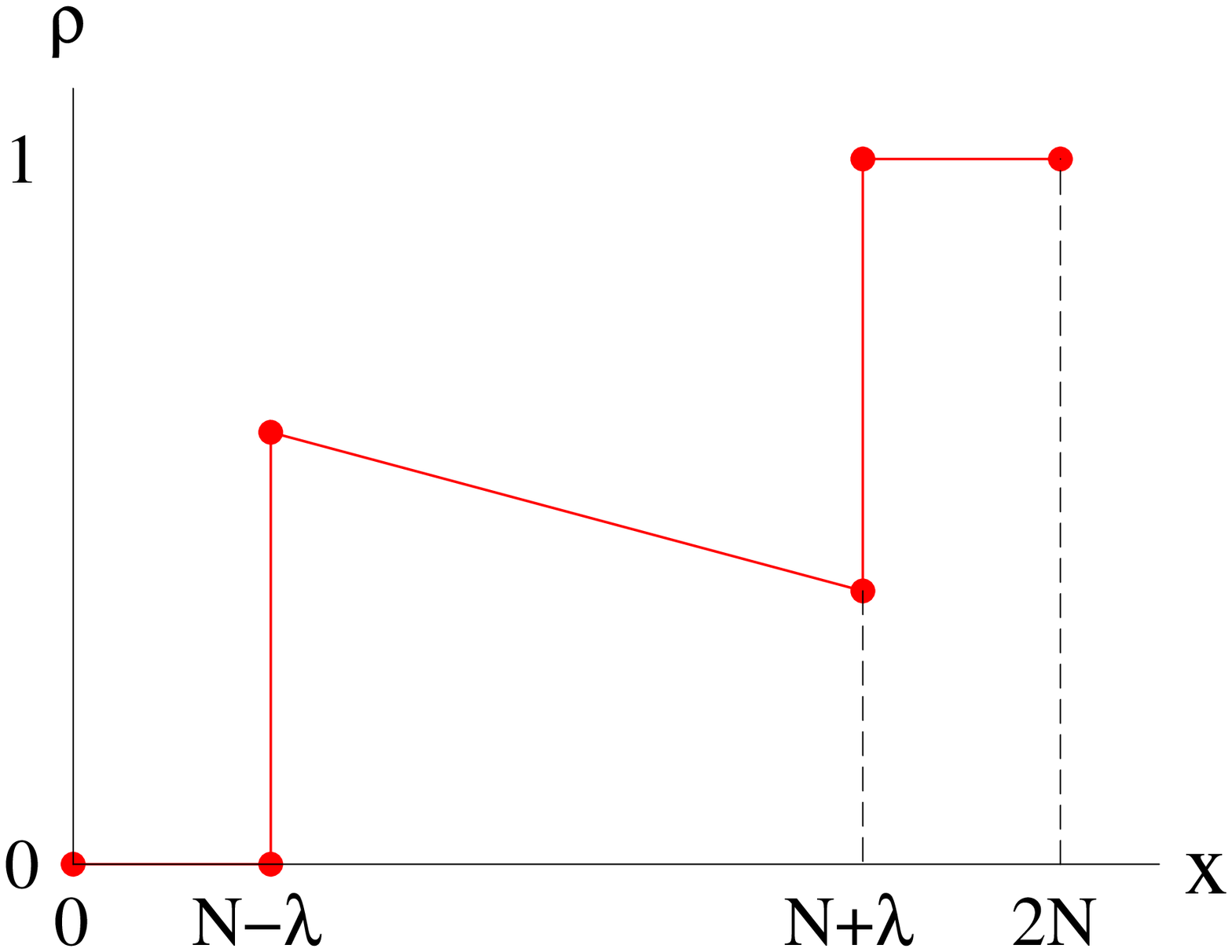}
\hskip 15pt
\includegraphics[angle=-90,width=.34\linewidth]{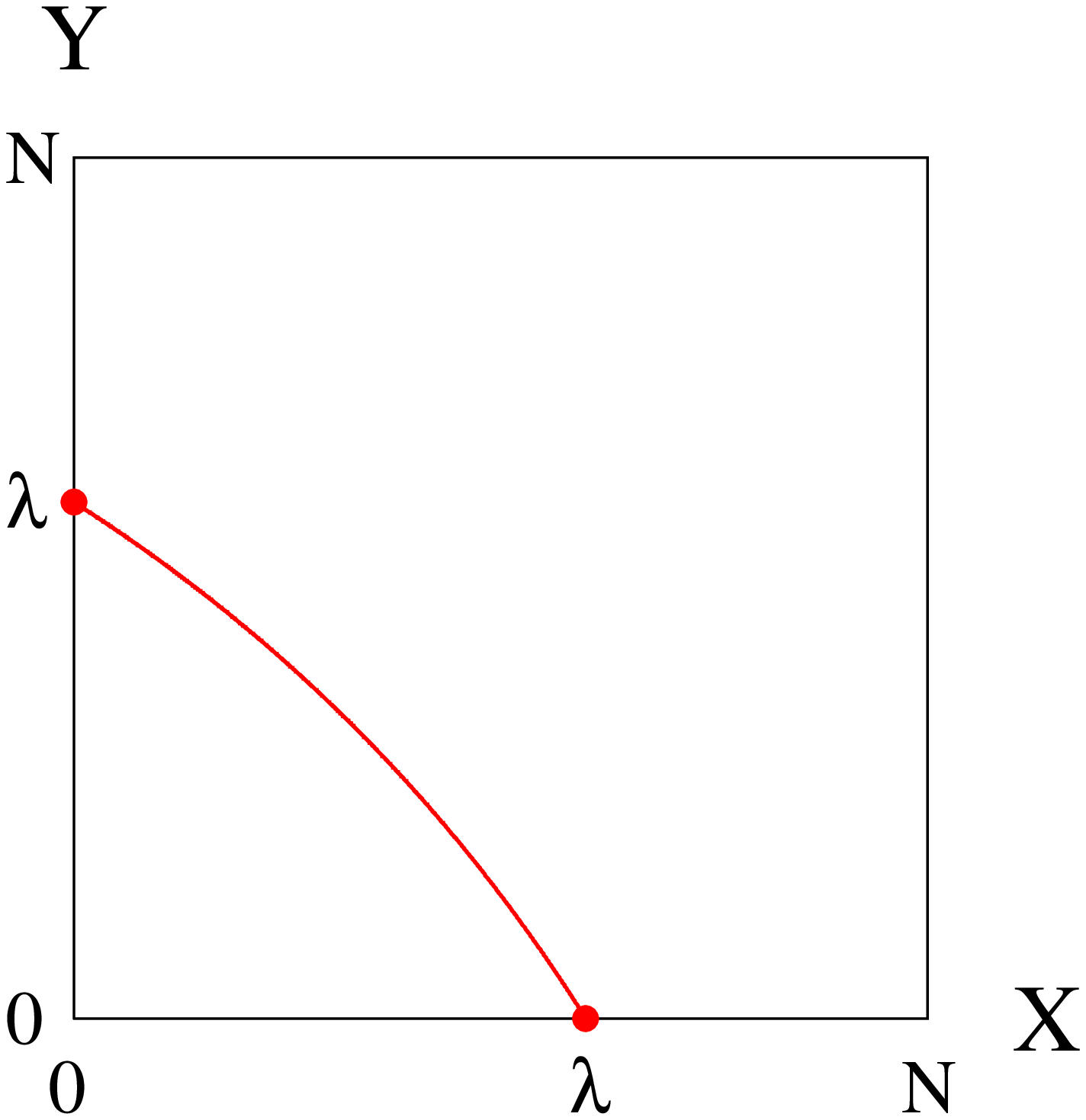}
\caption{\small
Stage II of zero-temperature dynamics for a square droplet.
Left: density profile of the TASEP.
Right: interface sweeping the droplet.}
\label{fig:s2}
\end{center}
\end{figure}

In this second stage, the finite system evolves differently from the infinite one.
Two shocks indeed invade the system from either end.
This corresponds to the fact that the West ($X=0$) and South ($Y=0$) sides of
the droplet begin to be eroded.
We have
\beq
\rho(x,t)=\left\{\matrix{
0\hfill&(0<x<N-\lambda),\hfill\cr
(N+t-x)/(2t)\quad&(N-\lambda<x<N+\lambda),\hfill\cr
1\hfill&(N+\lambda<x<2N)\hfill
\label{rho2}
}\right.
\eeq
(see figure~\ref{fig:s2}, left).
In the central region, the dynamics is the continuation of the previous stage.
The width $2\lambda$ of that active region is ruled by the differential equation
\beq
-\frac{\dd\lambda}{\dd t}=\frac{t-\lambda}{2t}.
\label{dlam}
\eeq
The expression on the right-hand side represents the velocity of the left shock,
given by the value of $1-\rho$ to its immediate right,
which can be read off from~(\ref{rho2}).
An integration of~(\ref{dlam}) with initial condition $\lambda=N$ for $t=N$ yields
\beq
\lambda=2\sqrt{Nt}-t.
\eeq
The active region shrinks in the course of time and disappears
at the end of the process, as $\lambda=0$ for $t=t^\star=4N$.
Throughout Stage~II, the shape of the interface is still given
by the parabolic arc~(\ref{para}).
Its apex still drifts ballistically according to~(\ref{apex}),
while its endpoints now sit on the West and South sides of the droplet
(see figure~\ref{fig:s2}, right).

\begin{figure}[!ht]
\begin{center}
\includegraphics[angle=-90,width=.45\linewidth]{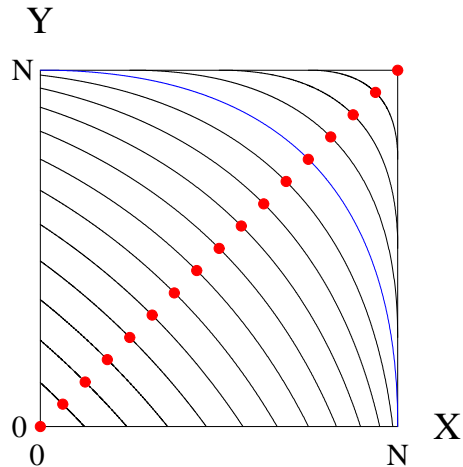}
\caption{\small
Shape of the interface sweeping a square droplet for zero-temperature dynamics
at 16 equally spaced times $t$ between 0 and $4N$.
Blue curve: shape of the interface at the end of Stage~I $(t=N$).
Red symbols: apex of the interface moving at constant velocity along the
diagonal,
according to~(\ref{apex}).}
\label{fig:carre}
\end{center}
\end{figure}

As a summary, figure~\ref{fig:carre} shows the successive interface shapes
at 16 equally spaced instants of time $t$ between 0 and $t^\star=4N$.
The blue curve shows the interface at the end of the first stage $(t=N=t^\star/4$),
when it touches the NW and SE corners of the droplet.
Even though the interface is macroscopically curved at all times,
its apex (red symbols) drifts along the diagonal with a constant velocity,
according to~(\ref{apex}).
This explains the simplicity of the ballistic law~(\ref{tstar}),
in spite of the non-triviality of the interface shape.

Near the end of the droplet's life ($t^\star-t\ll t^\star$),
its size is much smaller than its initial size,
and it has the shape of a right isosceles triangle.
This limiting triangular shape turns out to be universal,
in the sense that it is observed (at least) for all initial shapes
which are convex and symmetric with respect to the direction of the bias.
This universality has been noticed in several works devoted to generic coexistence
caused by spatial asymmetric dynamics~\cite{grinstein3,heringa,munoz},
as well as in~\cite{gp2015}.
The right isosceles triangular shape is indeed singled out
by the zero-temperature dynamics:
it is the only droplet shape which shrinks in a self-similar way,
i.e., keeping its shape, with its West and South sides being immobile,
and its hypotenuse moving at constant velocity~$v$.
Universal properties of limiting domain shapes have also been discussed
for other classes of kinetic Ising models~\cite{krt,karma,krap2,krap3,kms}.

\subsection{Fluctuations around the ballistic law}
\label{sec:fluctuat}

We now consider the fluctuations of the lifetime of individual droplets
around the ballistic law~(\ref{tstar}).
As discussed in section~\ref{sec:interface},
a square droplet of size $N$ corresponds to the TASEP on a closed interval of length $2N$,
and its lifetime $t^\star$ is the duration of the process~(\ref{reac}).
This lifetime is a random quantity
which depends on the whole stochastic history of the system.
In order to describe its fluctuations,
we have to go one step beyond the Burgers equation~(\ref{b}),
and to consider the Kardar-Parisi-Zhang (KPZ) theory
(see~\cite{hht} for an overview).
This approach predicts that the deterministic ballistic law~(\ref{tstar})
is complemented by a subleading fluctuating part of the form
\beq
t^\star\approx 4N+N^{1/3}\chi,
\label{chidef}
\eeq
where $\chi$ is a dimensionless random variable with a non-trivial limiting distribution.

Numerical simulations (see below) corroborate the following ansatz for the variable $\chi$:
\beq
\chi=2^{4/3}\xi_\gue,
\label{gue}
\eeq
where $\xi_\gue$ is distributed according to the Tracy-Widom $F_2$ law.
This distribution, originally derived
for the largest eigenvalue of a large Hermitian matrix in the Gaussian Unitary
Ensemble (GUE)~\cite{tw}, has then been shown to describe fluctuations
in a great deal of problems,
including the KPZ equation in various situations~\cite{j,psjsp,sh,acq}.
One of these cases is of particular relevance to the present context,
as it concerns the infinite TASEP with the step initial condition
where all sites to the left of the origin are occupied,
and all sites to the right are empty.
Johansson~\cite{j} has shown by rigorous means
that the time $t_N$ at which the $N$-th rightmost particle
arrives at the origin scales as
\beq
t_N\approx4N+2^{4/3}N^{1/3}\xi_\gue
\label{jo}
\eeq
for $N$ large enough.
An identical result has been obtained more recently~\cite{er}
for the interface propagating on a quantum chain of free fermions
prepared in a step initial condition.
A similar scaling law has been found for the anchored Toom interface~\cite{dlss}.
There, stationary height fluctuations obey a scaling law
analogous to~(\ref{chidef}) or~(\ref{jo}),
involving the variable $\xi_\goe$
distributed according to the Tracy-Widom~$F_1$ law,
characteristic of the Gaussian Orthogonal Ensemble (GOE)~\cite{bfls}.

The ansatz~(\ref{gue}) therefore amounts to stating
that the lifetime fluctuations of a large droplet are distributed
as if the full process~(\ref{reac}) were described by Stage~I.
During Stage~I, the active region of the system indeed behaves exactly
as an infinite TASEP launched from a step initial condition.
In this form, the above ansatz is made plausible by the observation,
already made in section~\ref{sec:interface},
that the dynamics in Stage~II is in several respects the continuation of Stage~I.
In particular, the same parabolic interface~(\ref{para}) holds throughout both stages.

\begin{figure}[!ht]
\begin{center}
\includegraphics[angle=-90,width=.5\linewidth]{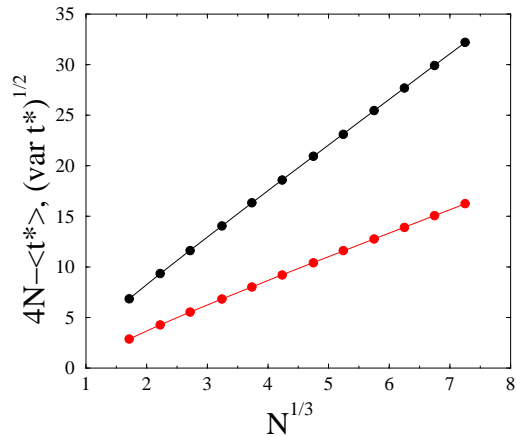}
\caption{\small
Statistics of the droplet lifetime $t^\star$ as measured on the TASEP,
plotted against $N^{1/3}$.
Upper (black) symbols: difference $4N-\mean{t^\star}$.
Lower (red) symbols: $(\var t^\star)^{1/2}$.
Full lines: fits of the form $AN^{1/3}+BN^{-1/3}$ yielding the
estimates~(\ref{moms}).}
\label{fig:chi}
\end{center}
\end{figure}

The above ansatz is strongly corroborated by the results
of numerical simulations on the TASEP.
Figure~\ref{fig:chi} shows plots of $4N-\mean{t^\star}$ and $(\var t^\star)^{1/2}$
against $N^{1/3}$ for TASEP sizes $2N$ ranging from 10 to 800.
Both datasets are rather well represented by linear fits of the form $aN^{1/3}+b$,
but fits of the form $AN^{1/3}+BN^{-1/3}$ are almost perfect.
The latter fits, shown as full lines, yield the estimates
\beq
\mean{\chi}\approx -4.47,\qquad\var\chi\approx 5.17
\label{moms}
\eeq
for the mean and variance of $\chi$.
These numbers are in perfect agreement with the predictions of the
ansatz~(\ref{gue}), i.e.~\cite{born},
\beq
2^{4/3}\mean{\xi_\gue}\approx-4.462859,\qquad2^{8/3}\,\var\xi_\gue\approx5.163465.
\eeq

\begin{figure}[!ht]
\begin{center}
\includegraphics[angle=-90,width=.5\linewidth]{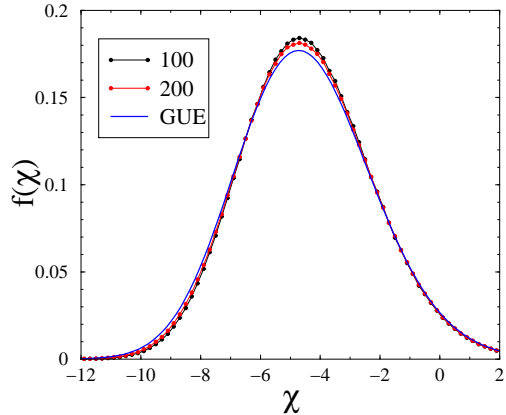}
\caption{\small
Distribution $f(\chi)$ of the variable $\chi$ introduced in~(\ref{chidef}),
as measured on the TASEP with $N=100$ and $N=200$.
Blue curve: limiting distribution predicted by the ansatz~(\ref{gue}).}
\label{fig:hchi}
\end{center}
\end{figure}

Figure~\ref{fig:hchi} shows a plot of the probability distribution $f(\chi)$
of the variable $\chi$,
as measured on the TASEP with $N=100$ and $N=200$.
Data are gathered over $10^7$ samples for each system size.
The distribution predicted by the ansatz~(\ref{gue}) is shown for comparison.
It reads $2^{-4/3}f_2(2^{-4/3}\chi)$,
where $f_2$ is the density of the Tracy-Widom~$F_2$ law.
A table of numerical values of $f_2$ is available at the web site~\cite{pswww}.
The very smooth convergence of the data to the proposed limiting form
provides a strong validation of the ansatz~(\ref{gue}).

\section{Finite-temperature zero-field dynamics}
\label{sec:finitet}

We now turn to the finite-temperature situation.
In this section we consider the dynamics in the absence of a magnetic field,
defined by the flipping rate~(\ref{eq:kun}),
while the finite-temperature dynamics in a field,
defined by the rate~(\ref{eq:wfield}), is studied in
section~\ref{sec:finiteth}.

Finite-temperature dynamics creates and annihilates bulk excitations.
Therefore,
at variance with the zero-temperature situation considered in
section~\ref{sec:zerot},
it is impossible to derive a closed effective dynamics for single interfaces.
We are thus led to study the full spin model by means of numerical simulations.
All data will be presented as a function of physical time $\alpha t$,
the time scale $\alpha$ being small enough that all rates are smaller than unity,
and can be used as flipping probabilities in a random sequential updating scheme.

\subsection{Overall picture}
\label{sec:overall}

\begin{figure}[!ht]
\begin{center}
\includegraphics[angle=-90,width=.3\linewidth]{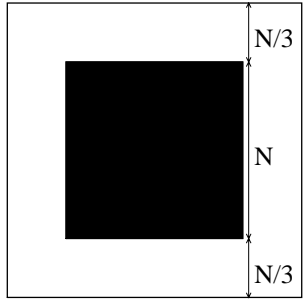}
\caption{\small
Initial condition used in numerical simulations at finite temperature.
A square droplet of size $N$ of down spins (filled)
is embedded in a square sea of size $N_\sea=5N/3$ of up spins,
with periodic boundary conditions along the sides of the outer frame.}
\label{fig:ile}
\end{center}
\end{figure}

We shall focus our attention on the fate of a large square droplet
of linear size~$N$ consisting initially only of down spins,
with $N$ multiple of 3, embedded in a square sea of size
\beq
N_\sea=\frac{5N}{3}
\label{nsea}
\eeq
of up spins, with periodic boundary conditions, as depicted in
figure~\ref{fig:ile}.
The initial mean magnetization is therefore
$M(0)=1-2(N/N_\sea)^2=7/25=0.28$.
This value is much smaller than the spontaneous magnetization $M_0$ (see~(\ref{spontan}))
for all temperatures considered hereafter in numerical simulations.

\begin{figure}[!ht]
\begin{center}
\includegraphics[angle=-90,width=.45\linewidth]{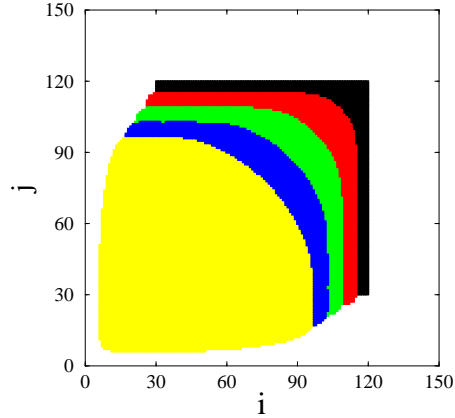}
\caption{\small
Average shape of a droplet of size $N=90$
embedded in a system of size $N_\sea=150$
for $T/T_c=0.5$,
at times $\alpha t$ multiples of 20.}
\label{fig:plot}
\end{center}
\end{figure}

Figure~\ref{fig:plot} shows the average shape of a droplet of size $N=90$
embedded in a sea of size $N_\sea=150$,
for $T/T_c=0.5$, at times $\alpha t$ multiples of~20.
The plotted shapes are obtained by averaging over 1000 thermal histories,
in order to practically eliminate any effect of thermal noise.
The dynamics of a large droplet concomitantly exhibits two distinct features:

\begin{itemize}
\item
It is advected ballistically in the SW direction,
i.e., along the direction of the bias.
Let $v_\cm$ be the velocity of this center-of-mass motion.
\item
It shrinks and disappears after a lifetime $t^\star$,
whose mean value also grows ballistically, i.e., linearly with $N$.
We denote by $v$ the associated {\it asymptotic shrinking velocity},
defined according to~(\ref{tsgen}).
In order to compare $v$ with its exactly known zero-temperature limit $v_0$
(see~(\ref{v0})), we rewrite~(\ref{tsgen}) as
\beq
\alpha\mean{t^\star}\approx\frac{2v_0}{v}N.
\label{vdef}
\eeq
\end{itemize}

\begin{figure}[!ht]
\begin{center}
\includegraphics[angle=-90,width=.45\linewidth]{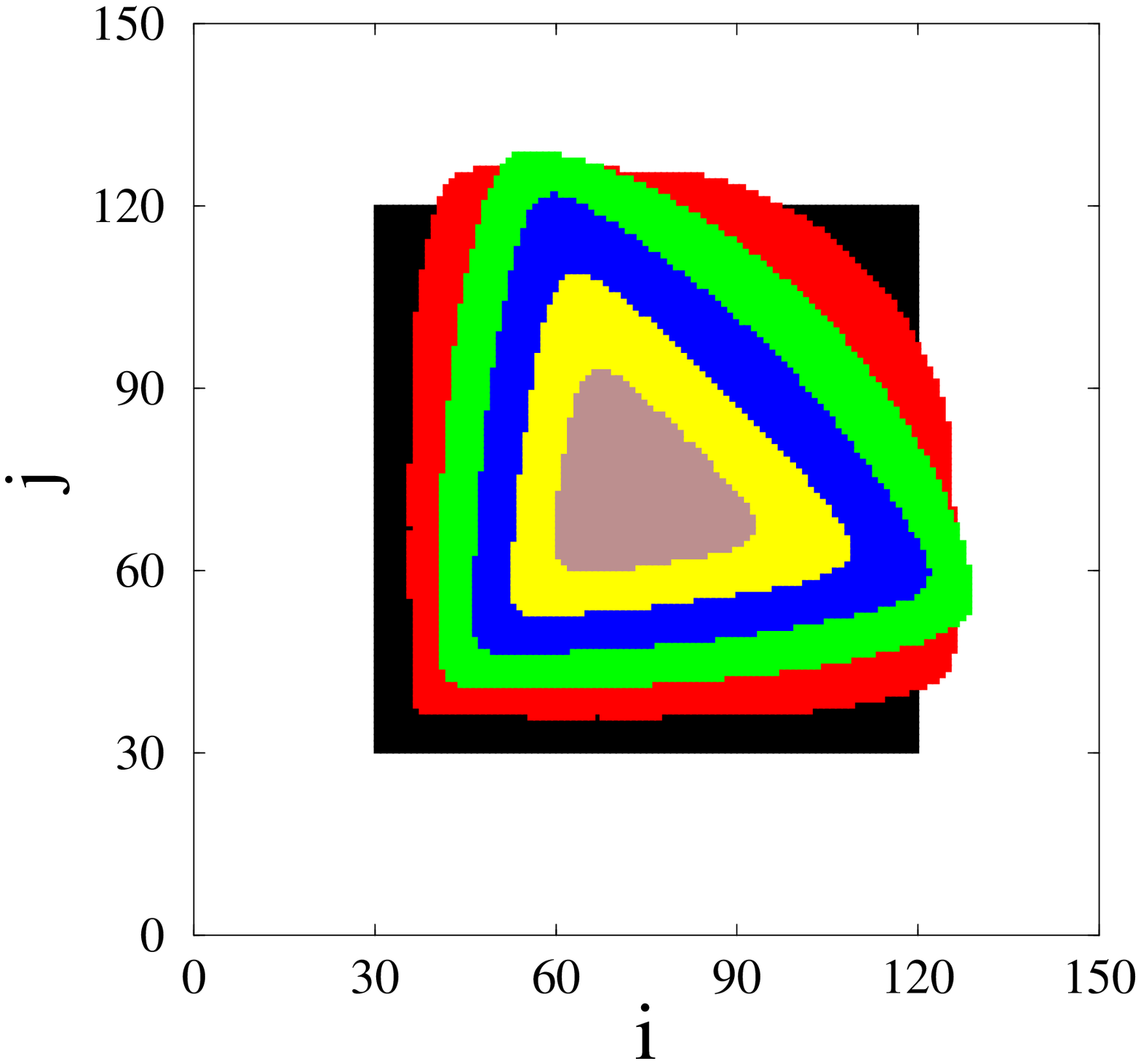}
\includegraphics[angle=-90,width=.45\linewidth]{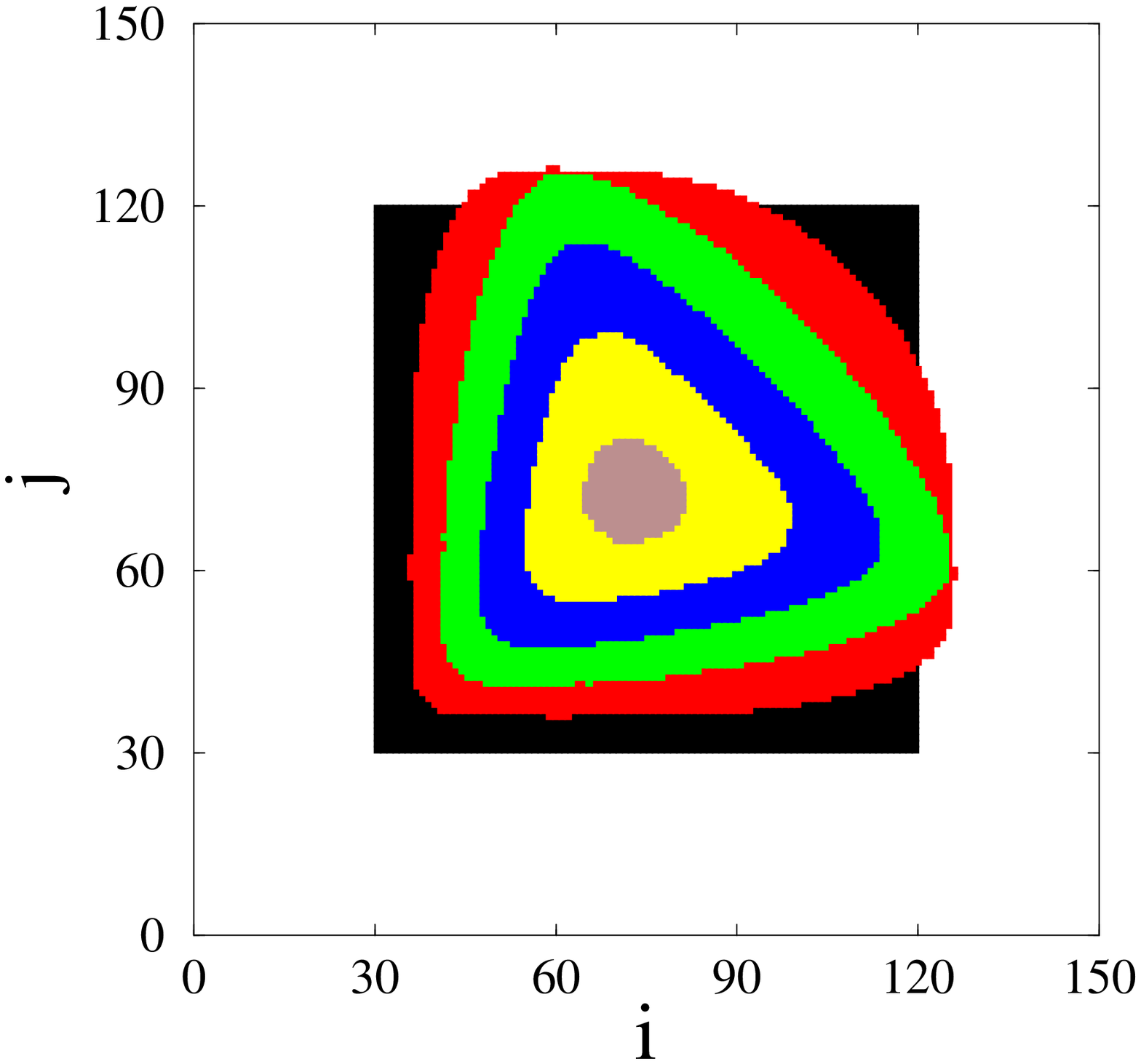}
\caption{\small
Back-advected average shapes of a droplet of size $N=90$.
Left: $T/T_c=0.5$, $v_\cm=0.78\,v_0$, times $\alpha t$ multiples of 60.
Right: $T/T_c=0.7$, $v_\cm=0.935\,v_0$, times $\alpha t$ multiples of 200.}
\label{fig:plots}
\end{center}
\end{figure}

In order to picture later stages of the dynamics,
it is convenient to advect back the droplet in the NE direction,
according to some estimated approximate value of~$v_\cm$,
so that the center of mass of the droplet remains essentially immobile
during its whole lifetime.
Figure~\ref{fig:plots} show such plots of a droplet of size $N=90$.
The left panel corresponds to $T/T_c=0.5$,
with $v_\cm\approx0.78\,v_0$ and times~$\alpha t$ multiples of 60.
The right panel corresponds to $T/T_c=0.7$,
with $v_\cm\approx0.935\,v_0$ and times $\alpha t$ multiples of~200.

As temperature increases from zero to the critical point,
the center-of-mass velocity $v_\cm$ increases steadily,
while the contours of the average droplet shape become progressively rounder,
and the mean droplet lifetime $\mean{t^\star}$ increases very fast.

\subsection{Asymptotic velocity}
\label{sec:finitev}

The goal of this section is to investigate the temperature dependence
of the asymptotic shrinking velocity $v$.
In the analysis of zero-temperature dynamics (see section~\ref{sec:zerot}),
a key role was played by the droplet lifetime $t^\star$.
At zero temperature, this lifetime is defined unambiguously
as the stopping time where the TASEP reaches the absorbing configuration
where all the particles are to the right of all the holes,
i.e., where the spin system reaches one of its ferromagnetic ground states.
In the present finite-temperature situation,
it will again be very convenient to measure a stopping time in numerical simulations.
The latter time must however be defined with care,
because the system keeps fluctuating.
We have chosen to define the stopping time $t^\star$ of a given run
as the first time where the mean magnetization per spin $M(t)$ exceeds (in
absolute value)
the spontaneous Onsager magnetization~$M_0$
of the infinite square lattice, given by~\cite{onsagermag}
\beq
M_0^8=1-\frac{1}{(\sinh(2/T))^4}.
\label{spontan}
\eeq
The stopping time $t^\star$ thus defined is a fluctuating quantity.
The linear growth of its mean value with $N$,
according to~(\ref{vdef}), allows us to measure the velocity $v$.

\begin{figure}[!ht]
\begin{center}
\includegraphics[angle=-90,width=.5\linewidth]{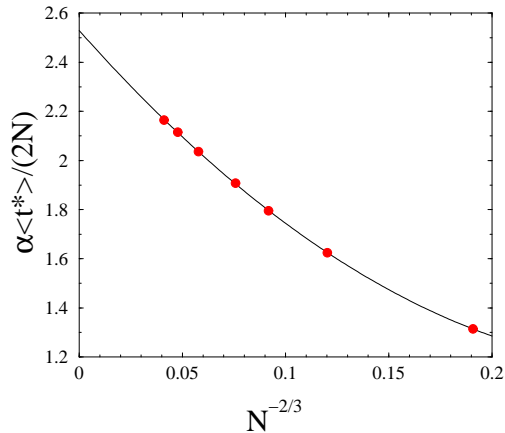}
\caption{\small
Ratio $\alpha\mean{t^\star}/(2N)$ against $N^{-2/3}$ for $T/T_c=0.5$.
Symbols: numerical data for $N$ ranging from 12 to 120.
Full curve: quadratic fit whose intercept yields $v_0/v\approx2.529$,
i.e., $v/v_0\approx0.395$.}
\label{fig:t5}
\end{center}
\end{figure}

Consider first the situation where $T/T_c=0.5$.
Figure~\ref{fig:t5} shows a plot of the ratio $\alpha\mean{t^\star}/(2N)$
against $N^{-2/3}$,
for $N$ ranging from 12 to 120, i.e., $N_\sea$~ranging from 20 to~200.
The choice of the abscissa axis is motivated by the zero-temperature
result~(\ref{chidef}),
implying that the first relative correction to the ballistic law
is proportional to $N^{-2/3}$.
The finite-temperature data are observed to exhibit the very same finite-size correction.
They are indeed well fitted by a quadratic function of $N^{-2/3}$,
shown as a full line, whose intercept yields $v_0/v\approx2.529$
(see~(\ref{vdef})), i.e., $v/v_0\approx0.395$.

\begin{figure}[!ht]
\begin{center}
\includegraphics[angle=-90,width=.5\linewidth]{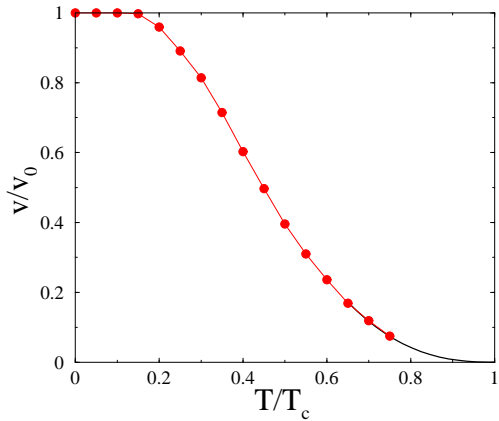}
\caption{\small
Ratio $v/v_0$ against reduced temperature $T/T_c$,
where $v$ is the velocity in zero field
and $v_0$ its known zero-temperature value (see~(\ref{v0})).
Symbols: numerical estimates.
Black line: power law~(\ref{vc}),
with exponent $\dv=2.5$ and amplitude fitted to data for $T/T_c>0.6$.}
\label{fig:vitesse}
\end{center}
\end{figure}

Repeating the above analysis for other values of temperature in the ordered phase,
we have obtained estimates for the ratio $v/v_0$
which are plotted against $T/T_c$ in figure~\ref{fig:vitesse}.
The above procedure yields accurate values of $v$ up to $T/T_c=0.75$.
At low temperature, $v$ is observed to depart from its zero-temperature value $v_0$ as
\beq
v\approx v_0(1-C\e^{-2/T}),
\eeq
with $C\approx4$.
A similar low-temperature correction in $\e^{-2/T}$
is shared by the free energy of an interface parallel to the axes,
\beq
\sigma=2+T\ln\tanh\frac{1}{T}\approx2(1-T\e^{-2/T}),
\eeq
another result due to Onsager~\cite{onsagerfree},
whereas for bulk thermodynamical quantities
the leading low-temperature correction is usually in $\e^{-4/T}$ (see~(\ref{uv})).
In the vicinity of the critical temperature,
the fast fall-off of the velocity is well described by the power law~(\ref{vc}),
shown as a full line.
The critical exponent $\dv\approx2.5$ will be determined
more accurately in section~\ref{sec:vcrit} by means of a finite-size scaling analysis.

\subsection{Critical region}
\label{sec:vcrit}

The critical region deserves a more careful analysis.
It is clear from figure~\ref{fig:vitesse}
that the velocity $v$ vanishes as $T_c$ is approached.
We expect that this slowing down is described by a power law
of the relative distance to the critical point,
\beq
\eps=\frac{T_c-T}{T_c},
\label{epsdef}
\eeq
and we denote by $\dv$ the corresponding critical exponent:
\beq
v\sim\eps^\dv.
\label{vc}
\eeq

At the critical temperature ($T=T_c$), the velocity $v$ vanishes,
and so the mean stopping time $\mean{t^\star}$
grows faster than ballistically with the droplet size $N$.
It is expected to diverge as
\beq
\mean{t^\star}\sim N^z,
\label{tc}
\eeq
where $z$ is the dynamical critical exponent of the model.
Numerical values of $z$ reported in~\cite{hcj}
for the anisotropic kinetic Ising model exhibiting generic coexistence studied
in~\cite{heringa}
suggest that the spatial anisotropy of the dynamics is irrelevant,
i.e., that $z$ coincides with the dynamical critical exponent
of the two-dimensional Ising model with conventional Glauber dynamics,
for which the more accurate estimate $z\approx2.167$ is available~\cite{bn}.

The power laws~(\ref{vc}) and~(\ref{tc})
can be embedded into the following finite-size scaling law:
\beq
\frac{N^z}{\alpha\mean{t^\star}}\approx F(x),\qquad x=\eps\,N^{(z-1)/\dv}.
\label{vfsslaw}
\eeq
The critical behavior~(\ref{tc}) of the stopping time
corresponds to $F(0)$ being some finite number,
while the critical slowing down~(\ref{vc}) of the velocity
corresponds to the power-law growth $F(x)\sim x^\dv$ for $x\gg1$.

\begin{figure}[!ht]
\begin{center}
\includegraphics[angle=-90,width=.5\linewidth]{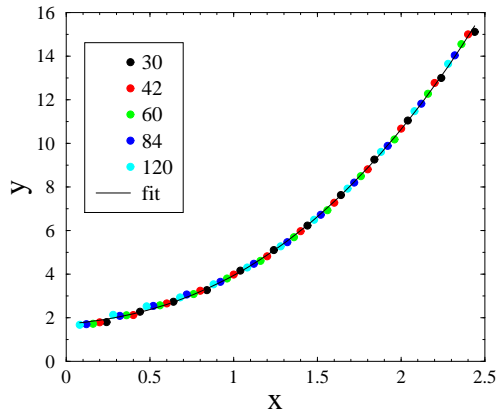}
\caption{\small
Critical finite-size scaling plot of the zero-field stopping time:
$y=N^z/(\alpha\mean{t^\star})$ is plotted against $x=\eps N^{(z-1)/\dv}$,
with $z=2.167$ and $\dv=2.5$.
Symbol colors denote values of $N$.
Full line: three-parameter fit (see text).}
\label{fig:vfss}
\end{center}
\end{figure}

In figure~\ref{fig:vfss} we have plotted
$y=N^z/(\alpha\mean{t^\star})$ against $x=\eps N^{(z-1)/\dv}$
for several~$N$ and~$\eps$,
in order to draw the finite-size scaling function $y=F(x)$.
Taking the value $z\approx2.167$ for granted,
a convincing data collapse is observed for $\dv=2.5$,
which has been chosen to draw the figure.
This value of~$\dv$ is corroborated by the good agreement between the data
and the three-parameter fit of the form $F(x)=a+b(x+c)^{2.5}$ (full line).
The latter fit is however more of a guide to the eye than an independent
determination of the exponent $\dv$.
We have checked that the above analysis is robust in several respects.
First, relaxing the hypothesis that $z\approx2.167$,
we consistently obtain a best data collapse for $z\approx2.2$.
Moreover, allowing a non-linear dependence of the scaling variable $x$
on the distance to the critical point,
i.e., changing~$\eps$ either to $\eps(1+A\eps)$ or to $\eps/(1+B\eps)$,
can possibly accommodate values of $\dv$ ranging from~2.3 to~2.7,
but hardly beyond these bounds.
We can therefore confidently give the estimate
\beq
\dv\approx2.5\pm0.2.
\label{dvres}
\eeq
We shall come back to this value of the critical exponent
at the end of section~\ref{sec:hcrit}.

\section{Finite-temperature dynamics in a field}
\label{sec:finiteth}

We now turn to the generic situation
where the temperature $T$ and the magnetic field~$h$ are non-zero,
so that the flipping rate is given by~(\ref{eq:wfield}),
with parameters $\gamma$ and~$\kappa$.
Our main goal is to obtain an accurate determination of the phase-boundary fields
$\pm\hb(T)$ delimiting the region of bistability (see figure~\ref{phase}),
especially in the low-temperature regime and in the vicinity of the critical point.

\subsection{Overall picture}
\label{sec:hover}

We again focus our attention on the fate of a large square droplet of down spins
embedded in a sea of up spins (see figure~\ref{fig:ile}).
The following picture is expected to hold:

\begin{itemize}
\item
For small enough values of the magnetic field ($\abs{h}<\hb(T)$),
the situation is qualitatively similar to what happens for $h=0$
(see section~\ref{sec:finitet}).
A large droplet disappears in a ballistic time.
The corresponding shrinking velocity $v$,
defined according to~(\ref{vdef}), now depends both on $T$ and~$h$.
It vanishes as $h\to-\hb(T)$, where the up phase loses its stability,
and therefore its ability to heal defects spontaneously.
The dynamics around $h=-\hb(T)$ is accordingly expected to be very slow.
\item
For large negative values of the magnetic field ($h<-\hb$),
the up phase of the sea has lost its stability,
and so the system ends up in a homogeneous down phase.
\item
For large positive values of the magnetic field ($h>\hb$),
the down phase inside the droplet has lost its stability,
and so the droplet may disappear faster than ballistically.
In any case the system ends up in a homogeneous up phase.
\end{itemize}

\subsection{Boundary field}
\label{sec:bdy}

The goal of this section is to study the temperature dependence
of the phase-boundary field~$\hb(T)$ delimiting the region of bistability.
Here again, it will be very convenient to measure a stopping time $t^\star$
by means of numerical simulations.
At variance with the situation in the absence of a field,
the magnetizations of the two nonequilibrium steady states
corresponding to the up and down phases are not known exactly.
This difficulty can be overcome by monitoring simultaneously
two samples of size $N_\sea$ with the same thermal noise,
albeit with two different initial conditions:
\begin{itemize}
\item[(1)]
a droplet of down spins of size $N$ in a sea of up spins,
as depicted in figure~\ref{fig:ile};
\item[(2)]
the uniform configuration where all spins are up.
\end{itemize}

We define the stopping time $t^\star$ of a given run as the first time
where the magnetization per spin $M^{(1)}(t)$ of the system with droplet
initial condition
exceeds in absolute value
the magnetization $M^{(2)}(t)$ of the initially homogeneous system.
The comparison of the absolute values of the two magnetizations
has the advantage of keeping the operational power of the chosen definition of
$t^\star$
for magnetic fields (slightly) below $-\hb$,
where the up phase has turned from stable to long-lived metastable.
We also introduce the {\it splitting probability} $p^\star$~\cite{kampen}
of the up phase,
defined as the probability for the system with droplet initial condition to end
in the up phase, i.e., in practice, to have $M^{(1)}(t^\star)>0$.

\begin{figure}[!ht]
\begin{center}
\includegraphics[angle=-90,width=.45\linewidth]{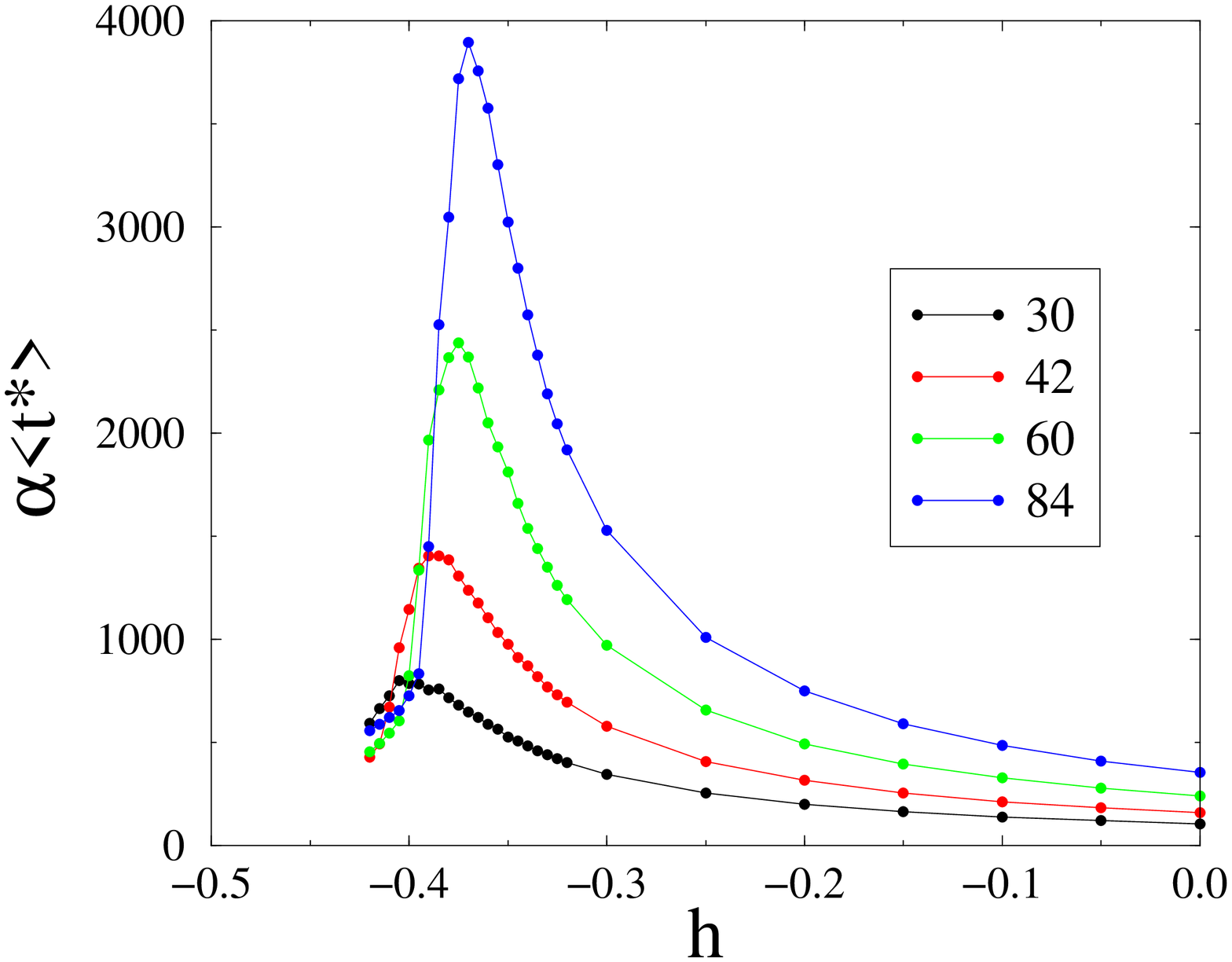}
\includegraphics[angle=-90,width=.45\linewidth]{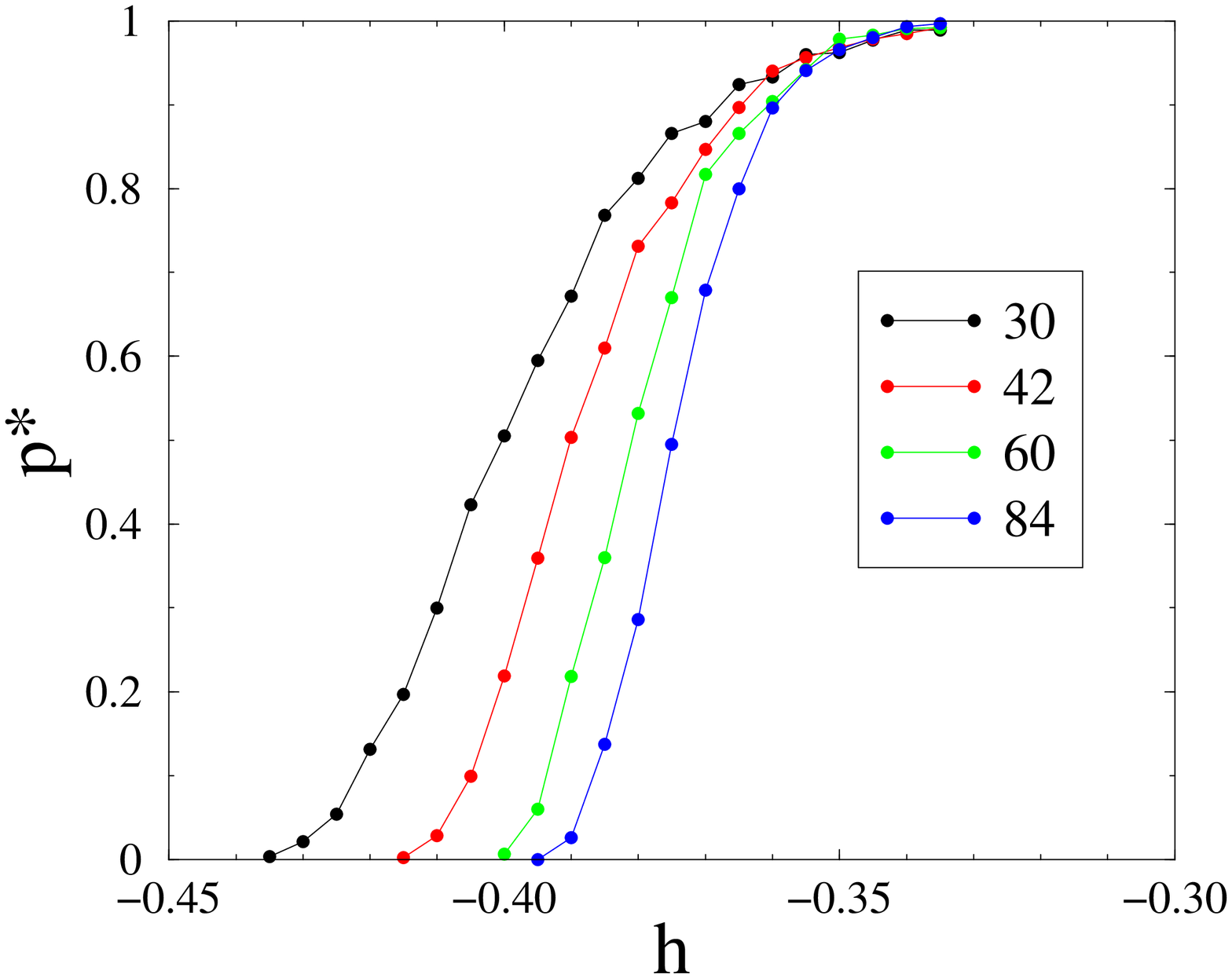}
\caption{\small
Mean stopping time $\alpha\mean{t^\star}$ (left)
and splitting probability $p^\star$ (right),
against magnetic field $h$ for $T/T_c=0.5$.
Symbol colors denote values of $N$.}
\label{fig:cinq}
\end{center}
\end{figure}

Consider first the situation where $T/T_c=0.5$.
Figure~\ref{fig:cinq} shows
the mean stopping time $\alpha\mean{t^\star}$ (left)
and of the splitting probability~$p^\star$ (right),
as functions of the magnetic field $h$, for several droplet sizes $N$.
As $N$ increases,
the stopping time exhibits peaks which get higher and narrower,
whereas the splitting probability crosses over more and more steeply from 0 to 1.
These observations are fully consistent with the overall picture
of section~\ref{sec:hover}.
For a given droplet size $N$,
the value $h^{(N)}$ of the magnetic field where the stopping time is maximal
is virtually identical with the middle point where $p^\star=1/2$.
The values $h^{(N)}$ thus defined shift slowly to the right as $N$ increases,
toward a limit which is nothing but $-\hb(T)$.
This yields $\hb\approx0.35$ for $T/T_c=0.5$.

\begin{figure}[!ht]
\begin{center}
\includegraphics[angle=-90,width=.5\linewidth]{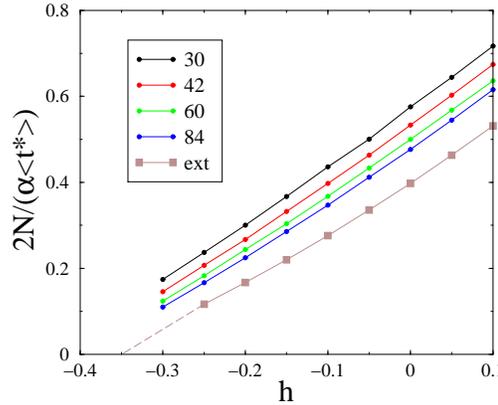}
\caption{\small
Ratio $2N/(\alpha\mean{t^\star})$ against $h$ for $T/T_c=0.5$.
Symbol colors denote values of $N$.
Brown squares: extrapolated values yielding $v/v_0$ as a function of $h$.
The dashed straight line suggests that $v$ vanishes linearly with $h+\hb$,
with $\hb\approx0.35$.}
\label{fig:vcinq}
\end{center}
\end{figure}

In order to look at the problem from a different perspective,
we have measured the dependence of the velocity $v$ on the magnetic field $h$.
To do so, data for $\alpha\mean{t^\star}$ have been extrapolated,
along the lines of section~\ref{sec:finitev},
by means of quadratic fits in the variable $N^{-2/3}$.
Figure~\ref{fig:vcinq} shows plots of the ratio $2N/(\alpha\mean{t^\star})$
against $h$ for several~$N$,
as well as the extrapolated values yielding a plot of the ratio $v/v_0$ against~$h$.
The dashed straight line suggests that $v/v_0$ vanishes linearly
with the distance $h+\hb$ to the boundary of the region of bistability.
We thus consistently recover $\hb\approx0.35$ for $T/T_c=0.5$.

\begin{figure}[!ht]
\begin{center}
\includegraphics[angle=-90,width=.5\linewidth]{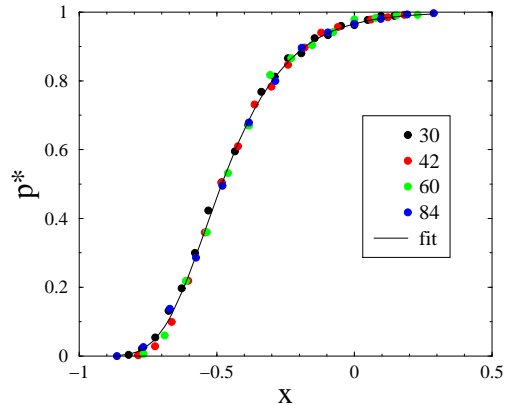}
\caption{\small
Finite-size scaling plot of the splitting probability $p^\star$
against $x=N^{2/3}(h+\hb)$ for $T/T_c=0.5$, with $\hb=0.35$.
Symbol colors denote values of $N$.
The sigmoidal fit is a guide to the eye.}
\label{fig:pfss}
\end{center}
\end{figure}

Finally, the splitting probability $p^\star$
is observed to obey a finite-size scaling law of the form
\beq
p^\star\approx\Phi(x),\qquad x=N^{2/3}(h+\hb),
\label{pfsslaw}
\eeq
near the phase-boundary field $-\hb$,
where the crossover exponent 2/3 is again suggested
by the relative size of the fluctuating term in~(\ref{chidef}),
stemming from the KPZ theory.
The data of figure~\ref{fig:cinq} for $p^\star$
are replotted against $x=N^{2/3}(h+\hb)$ in figure~\ref{fig:pfss},
where the value $\hb=0.35$ has been chosen in order to provide the best data collapse.
This value of the phase-boundary field agrees with the estimates
drawn from the observations made above on figures~\ref{fig:cinq} and~\ref{fig:vcinq}.
This finite-size scaling method is however more accurate,
as the estimated uncertainty on $\hb$ is smaller than~0.005.
The sigmoidal fit shown as a full line is meant as a guide to the eye.

The finite-size scaling ansatz~(\ref{pfsslaw})
explains the slow convergence of the $h^{(N)}$ observed in figure~\ref{fig:cinq},
as it predicts $h^{(N)}+\hb\sim N^{-2/3}$.
The corresponding maxima of the stopping time
are observed to obey an effective power law of the form $t^\star_\max\sim N^{1.2}$,
with an apparent dynamical exponent much smaller than the exponent $z\approx2.167$
entering the zero-field critical law~(\ref{tc}).
In any case, the dependence of $t^\star_\max$
on the system size $N$ is expected to be more complex than a single power law.
A good deal of works~\cite{hck,rtm,thc,lfg,svb},
elaborating on the well-known droplet nucleation theory~\cite{f,l},
have indeed revealed that several time scales
enter the dynamics of metastable phases in finite systems.

\begin{figure}[!ht]
\begin{center}
\includegraphics[angle=-90,width=.5\linewidth]{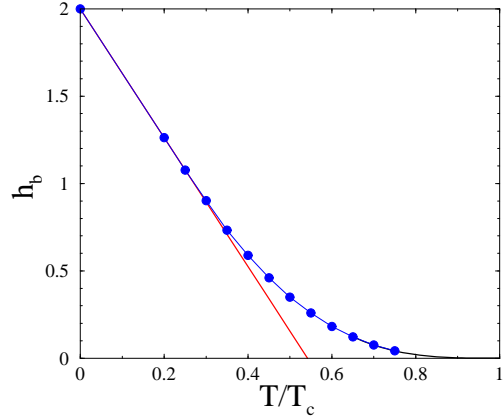}
\caption{\small
Boundary field $\hb$ against reduced temperature $T/T_c$.
Symbols: numerical estimates (see text).
Red straight line: fit $\hb=2-BT$ of the low-temperature behavior, with $B=1.63$.
Black line: power law~(\ref{hc}),
with exponent $\dh=3.2$ and amplitude fitted to data for $T/T_c>0.6$.}
\label{fig:h0}
\end{center}
\end{figure}

The finite-size scaling analysis of the splitting probability,
yielding the most accurate determination of the phase-boundary field $\hb(T)$,
has been repeated for other values of temperature in the ordered phase.
This procedure yields accurate values of $\hb$ up to $T/T_c=0.75$.
Figure~\ref{fig:h0} shows a plot of the estimates
thus obtained against $T/T_c$, providing a quantitative version
of the phase diagram shown schematically in figure~\ref{phase}.
At low temperature, the phase-boundary field is observed to behave linearly~as
\beq
\hb\approx2-BT,
\label{hb}
\eeq
with $B\approx1.63$, with the next correction being presumably exponentially small.
The rationale behind this formula is that
the two low-temperature expansion variables
\beq
a=\e^{-4/T},\qquad b=\e^{-2h/T},
\label{uv}
\eeq
are proportional to each other along the phase boundary line,
as $b\approx qa$ translates to~(\ref{hb}) with $B=(\ln q)/2$.
The fast fall-off of the phase-boundary field
as $T_c$ is approached is described by the power law~(\ref{hc}).

\subsection{Critical region}
\label{sec:hcrit}

\begin{figure}[!ht]
\begin{center}
\includegraphics[angle=-90,width=.5\linewidth]{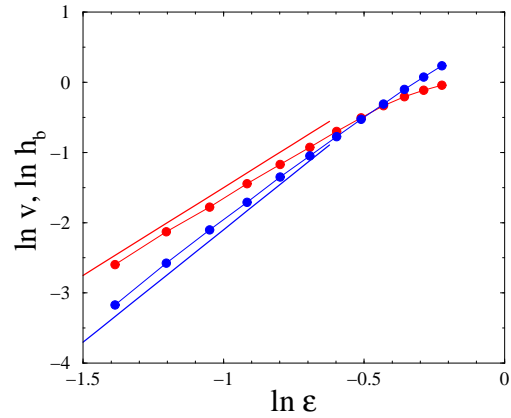}
\caption{\small
Logarithmic plots of the zero-field velocity $v$ (red)
and of the phase-boundary field $\hb$ (blue)
against $\eps=(T_c-T)/T_c$.
Symbols: data from figures~\ref{fig:vitesse} and~\ref{fig:h0}.
The slopes $\dv=2.5$ and $\dh=3.2$ of the straight lines
are taken from the outcomes~(\ref{dvres}) and~(\ref{dhres})
of more accurate finite-size scaling analyses.}
\label{fig:dvdh}
\end{center}
\end{figure}

The critical region again deserves a more careful analysis.
As observed long ago~\cite{grinstein1,heringa},
the phase-boundary field $\hb$ vanishes as a power law as $T_c$ is approached, as
\beq
\hb\sim\eps^\dh.
\label{hc}
\eeq
The critical exponent $\dh$ is larger than its counterpart $\dv$ by a sizeable amount.
This is evidenced in figure~\ref{fig:dvdh},
showing logarithmic plots of the zero-field velocity~$v$ and the phase-boundary field $\hb$
(data taken from figures~\ref{fig:vitesse} and~\ref{fig:h0}) against $\eps$.
The data are observed to obey two different power laws.
The slopes $\dv=2.5$ and $\dh=3.2$ of the straight lines
are the outcomes~(\ref{dvres}) and~(\ref{dhres})
of more accurate finite-size scaling analyses.
Least-square fits to the three to five leftmost data points
however essentially yield the same exponents, especially so for $\dh$.

In order to measure the exponent $\dh$ with good accuracy,
we again make use of finite-size scaling theory.
An appropriate observable is provided by the
slope at $h=0$ of the data plotted in figure~\ref{fig:vcinq}, i.e.,
\beq
\mu=\frac{2N}{\alpha}
\left(\frac{\partial}{\partial h}\frac{1}{\mean{t^\star}}\right)_{h=0}.
\eeq
This quantity becomes the reduced mobility
\beq
\mu=\frac{1}{v_0}\left(\frac{\partial v}{\partial h}\right)_{h=0}
\eeq
in the thermodynamic limit.
It has several advantages from a numerical viewpoint:
it is very sensitive to a small magnetic field,
and can be measured accurately for droplet sizes up to $N=120$.
The nearly linear behavior of the data shown in figure~\ref{fig:vcinq}
suggests that $\mu$ scales as the ratio $v/\hb$,
and therefore that it diverges as
\beq
\mu\sim\eps^{-(\dh-\dv)}
\label{muc}
\eeq
as the critical is approached after taking the limit of a very large system.
The mobility~$\mu$ therefore gives direct access to the exponent difference $\dh-\dv$.

\begin{figure}[!ht]
\begin{center}
\includegraphics[angle=-90,width=.5\linewidth]{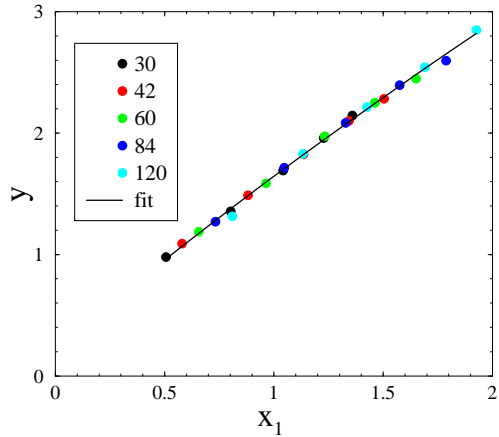}
\caption{\small
Critical finite-size scaling plot of the reduced mobility:
$y=N^{(z-1)(\dh-\dv)/\dv}/\mu$
is plotted against $x_1=\eps(1-A\eps)N^{(z-1)/\dv}$, with $z=2.167$,
$\dv=2.5$, $A=0.84$ and $\dh-\dv=0.7$.
Symbol colors denote values of $N$.
Full line: three-parameter fit (see text).}
\label{fig:mufss}
\end{center}
\end{figure}

The power law~(\ref{muc}) extends to the finite-size scaling law
\beq
\mu\approx N^{(z-1)(\dh-\dv)/\dv}G(x),
\label{mufsslaw}
\eeq
where $x=\eps\,N^{(z-1)/\dv}$ is as in~(\ref{vfsslaw}),
and the scaling function falls off as $G(x)\sim x^{-(\dh-\dv)}$ as $x\gg1$.
This finite-size scaling law is demonstrated by figure~\ref{fig:mufss},
where $y=N^{(z-1)(\dh-\dv)/\dv}/\mu$ is plotted
against $x_1=(1-A\eps)x=\eps(1-A\eps)N^{(z-1)/\dv}$.
In the definition of $x_1$,
$\eps$ has been changed to $\eps(1-A\eps)$
in order to empirically improve convergence.
Taking $z\approx2.167$ and $\dv\approx2.5$ for granted,
the best data collapse is observed for $A\approx0.84$ and $\dh-\dv\approx0.7$.
The latter value is consistent with the rather good agreement between the data
and the three-parameter fit of the form $y=1/G(x)=a+b(x+c)^{0.7}$ (full line).
Here again, the latter fit is more of a guide to the eye than an independent
determination of the exponent difference $\dh-\dv$.

We can state the outcome of the above analysis as
\beq
\dh-\dv\approx0.7\pm0.1.
\label{dhdvres}
\eeq
Combining this with $\dv\approx2.5\pm0.2$ (see~(\ref{dvres})), we obtain
finally
\beq
\dh\approx3.2\pm0.3.
\label{dhres}
\eeq

The above predictions can be compared with various results in the literature.
On the numerical side,
we know of no previous measurement of~$\dv$,
and of only two estimates of $\dh$,
namely $\dh\approx3.0\pm0.4$~\cite{grinstein1},
and $\dh\approx2.7\pm0.2$~\cite{heringa}.
Our estimate~(\ref{dhres}) agrees well with the first of those results,
albeit only marginally with the second.
It should be noticed that the error bar given in~(\ref{dhres}) is not reduced
with respect to those given in earlier works,
in spite of the good apparent quality of the finite-size scaling plots
of figures~\ref{fig:vfss} and~\ref{fig:mufss}.

It is also worth putting our results for the critical exponents $\dh$ and $\dv$
in perspective with a heuristic renormalization-group approach
put forward in~\cite{heringa}.
This line of thought yields
\beq
\dh=\frac{y_h+\abs{y_i}}{y_t}=\frac{15}{8}+\abs{y_i},
\label{dhrg}
\eeq
where $y_h=2\nu-\beta=\beta+\gamma=\beta\delta=15/8$ and $y_t=1/\nu=1$
are the usual scaling dimensions of the two-dimensional Ising model,
whereas $y_i<0$ is the scaling dimension
of some irrelevant operator left undetermined in~\cite{heringa}.
The same line of thought yields
\beq
\dv=z-1+\frac{\abs{y_i}}{y_t}\approx1.167+\abs{y_i},
\label{dvrg}
\eeq
taking again $z\approx2.167$ for granted,
and so the exponent difference is predicted to be
\beq
\dh-\dv=\frac{y_h}{y_t}+1-z\approx0.708,
\label{dhdvrg}
\eeq
irrespective of the scaling dimension $y_i$.
Our numerical result~(\ref{dhdvres}) is in perfect agreement with this prediction.
Moreover,~(\ref{dvres}) yields the estimate
\beq
y_i\approx-1.3\pm0.2
\label{yi}
\eeq
for the irrelevant scaling dimension entering the heuristic approach of~\cite{heringa}.

\subsection{Magnetization curves}
\label{sec:mag}

\begin{figure}[!ht]
\begin{center}
\includegraphics[angle=-90,width=.5\linewidth]{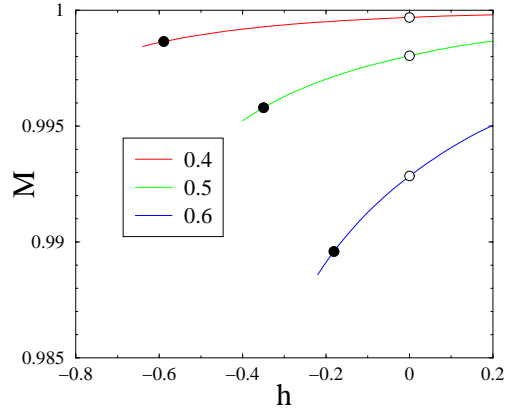}
\caption{\small
Mean magnetization $M$ of the up phase against magnetic field $h$,
for three temperatures in the ordered phase.
Colors denote values of $T/T_c$.
Open symbols: spontaneous magnetization $M_0$ at $h=0$ (see~(\ref{spontan})).
Full symbols: phase-boundary field $-\hb(T)$ and corresponding magnetization
$\mb(T)$.}
\label{fig:aim}
\end{center}
\end{figure}

To close our numerical investigation,
we show in figure~\ref{fig:aim} magnetization curves of the model
for three temperatures in the ordered phase ($T/T_c=0.4$, 0.5 and~0.6).
For each temperature, the mean magnetization $M$ of the nonequilibrium steady state
of the up phase is plotted against the magnetic field $h$.
Data are taken for samples of size $N_\sea=200$,
with uniform initial condition (2),
so that size effects are entirely negligible.
The spontaneous magnetization $M_0(T)$ (see~(\ref{spontan}))
is recovered at $h=0$ (open symbols), as should be.
Full symbols show the values of the phase-boundary field $-\hb(T)$
and of the corresponding magnetization $\mb(T)$.
From a merely numerical perspective,
the magnetization curves can easily be continued slightly below $-\hb(T)$,
where the up phase has turned from stable to metastable but very long-lived.

The magnetization curves are observed to be very regular near the phase-boundary field.
This corroborates the commonly accepted expectation
that nothing dramatic occurs
to the pure phases near the limits of their region of stability.

\ack

It is pleasure to thank Paul Krapivsky, David Mukamel, Tomohiro Sasamoto
and Gunter Sch\"utz for interesting discussions.

\end{document}